\documentclass[prd,aps,floats,epsfig,eqsecnum,nofootinbib]{revtex4}
\usepackage{amsmath,amssymb,verbatim,epsfig}
\begin{document}
\date{}
\title{Large scale magnetogenesis  from a non-equilibrium phase transition\\
in the radiation dominated era}
\author{D. Boyanovsky$^{(a,b)}$}
\email{boyan@pitt.edu}
\author{H. J. de Vega$^{(b,a)}$}
\email{devega@lpthe.jussieu.fr}
\author{M. Simionato $^{(a)}$}
\email{mis6@pitt.edu} 
\affiliation{$^{(a)}$ Department of Physics
and Astronomy, University of Pittsburgh, Pittsburgh, Pennsylvania 15260, USA\\
$^{(b)}$ LPTHE, Universit\'e Pierre et Marie Curie (Paris VI) et
Denis Diderot (Paris VII), Tour 16, 1er. \'etage, 4, Place
Jussieu, 75252 Paris, Cedex 05, France}

\begin{abstract}
We study the generation of large scale primordial magnetic fields
by a cosmological phase transition during the radiation dominated
era. The setting is a theory of $N$ charged scalar fields coupled
to an abelian gauge field, that undergoes a phase transition at a
critical temperature much larger than the electroweak scale. The
dynamics after the transition features two distinct stages: a
spinodal regime dominated by linear long-wavelength instabilities,
and a scaling stage in which the non-linearities and backreaction
of the scalar fields are dominant. This second stage describes the
growth of horizon sized domains. We implement a recently
introduced formulation to obtain the spectrum of magnetic fields
that includes the dissipative effects of the plasma. We find that
large scale magnetogenesis is very efficient during the scaling
regime. The ratio between the energy density on scales larger than
$L$ and that in the background radiation $r(L,T)=
\rho_B(L,T)/\rho_{cmb}(T)$ is $r(L,T) \sim 10^{-34}$ at the
Electroweak scale and $r(L,T)\sim 10^{-14}$ at the QCD scale for
$L \sim 1~\mbox{Mpc}$. The resulting spectrum is insensitive to
the magnetic diffusion length. We conjecture that a similar
mechanism could be operative after the QCD chiral phase
transition.
\end{abstract}
\date{\today}

\maketitle

\tableofcontents 

\section{Introduction}
A variety of astrophysical observations including Zeeman
splitting, synchrotron emission, Faraday rotation measurements
(RM) combined with pulsar dispersion measurements (DM) and
polarization measurements suggest the presence of large scale
magnetic fields\cite{Parker,Kronberg,grasso,widrow,giova1,han}.
The  strength of typical galactic magnetic fields is measured to
be $\sim \mu~G$\cite{Kronberg,grasso,widrow,han} and they are
correlated on very large scales up to galactic or even larger
reaching to scales of cluster of galaxies $\sim
1~\mbox{Mpc}$\cite{Kronberg,grasso,widrow,giova1}. The origin of
these large scale magnetic fields is still a subject of much
discussion and controversy. It is currently agreed that a variety
of dynamo mechanisms are efficient in \emph{amplifying} seed
magnetic fields with typical growth rates $\Gamma \sim
\mbox{Gyr}^{-1}$ over time scales $\sim 10-12 $ Gyr (for a
thorough discussion of the mechanisms and  models
see\cite{widrow}). The ratio of the energy density of the seed
magnetic fields  on scales larger than $L$ (today) to that in the
cosmic background radiation, $r(L)=\rho_B(L)/\rho_{cmb}$ must be
$r(L\sim 1\mbox{Mpc})\geq 10^{-34}$ for a dynamo mechanism to
amplify it to the observed value, or $r(L\sim 1\mbox{Mpc})\geq
10^{-8}$ for the seed to be amplified solely by the gravitational
collapse of a protogalaxy\cite{giova1}.

There are also different  proposals to explain the origin of the
initial seed. Astrophysical batteries rely on gradients of the
charge density concentration and pressure and their efficiency in
producing seeds of the necessary amplitude is still very much
discussed\cite{Kronberg,widrow}. Primordial magnetic fields that
could be the seeds for dynamo amplification can be generated at
different stages in the history of the early Universe, in
particular during inflation, preheating and or phase
transitions\cite{grasso,giova1,widrow}. Primordial (hyper)
magnetic fields may have important consequences in electroweak
baryogenesis\cite{grassoEW}, Big Bang nucleosynthesis
(see\cite{grasso}), the polarization of the CMB\cite{durrer1} via
the same physical processes as Faraday rotation, and structure
formation\cite{grasso,giova1,dolgov2}, thus sparking an intense
program to study the origin and consequences of the generation of
magnetic fields in the early Universe\cite{hogan}-\cite{ahonen}.

A reliable estimate of the amplitude and correlations of seed
magnetic fields must include the dissipative properties of the
plasma, in particular the
conductivity\cite{turnerwidrow,Giovanni,giovashapo}. In
ref.\cite{magfiI} we have  introduced a formulation that allows to
compute the generation of magnetic fields from processes strongly
out of equilibrium. This formulation, which is based on the exact
set of Schwinger-Dyson equations for the transverse photon
propagator is manifestly gauge invariant and is general for any
matter fields and any cosmological background (conformally related
to Minkowski space-time). In the case in which strongly out of
equilibrium effects arise from long-wavelength fluctuations, such
as during phase transitions, this formulation allows to separate
the contribution of the hard degrees of freedom which are in local
thermodynamic equilibrium from that of the soft degrees of freedom
that fall out of LTE (local thermal equilibrium) during the phase
transition and whose
dynamics is strongly out of equilibrium. This separation of
degrees of freedom leads to a consistent incorporation of the
dissipative effects via the conductivity (for details
see\cite{magfiI}). In that reference a  study of magnetogenesis in
Minkowski space-time during a supercooled phase transitions was
presented and the results highlighted the main aspects of the
generation of magnetic and electric fields in these situations.

{\bf The goals of this article:} In this article we study the
generation of large scale (hyper) magnetic fields by a
cosmological phase transition during a radiation dominated era by
implementing the formulation introduced in ref.\cite{magfiI}. The
setting is a theory of $N$ charged scalar fields coupled to an
abelian gauge field (hypercharge). We consider the situation when
this theory undergoes a phase transition after the reheating stage
and before either the Electroweak or the QCD phase transition,
since we expect that these transitions will lead to new physical
phenomena. The non-perturbative dynamics out of equilibrium is
studied in the limit of a large number $N$ of (hyper) charged
fields and to leading order in the gauge coupling. The
non-equilibrium dynamics of the charged scalar sector features two
distinct stages. The first one describes the early and
intermediate time regime and is dominated by the spinodal
instabilities which are the hallmark of the process of phase
separation and domain formation and growth. This stage describes
the dynamics between the time at which the phase transition takes
place and that at which non-linearities become important via the
backreaction. The second stage corresponds to a \emph{scaling
regime} which describes the slower non-equilibrium evolution of
Goldstone bosons and the process of phase ordering\cite{scaling}
and growth of horizon-sized domains. This scaling regime is akin
to the solution found in the \emph{classical} evolution of scalar
field models with broken continuous symmetries after the phase
transition that form the basis for models of structure formation
based on topological defects\cite{turok,durrer}.

The solution of the scalar field dynamics \cite{scaling} is the
input in the expression for the spectrum of the magnetic field
obtained in~\cite{magfiI} to obtain the amplitude of the
primordial seed generated during both stages.

We find that scaling stage is the most important  for the
generation of large scale magnetic fields. Large scale magnetic
fields are generated via loop effects from the dynamics of modes
that are at the scale of the horizon or smaller. The resulting
spectrum is rather insensitive to the diffusion length scale which
is much smaller than the horizon during the radiation dominated
era. The ratio of the magnetic energy density on scales larger
than $L$ (today) to the energy density in the
background radiation $r(L,\eta)=\rho_B(L,\eta)/\rho_{cmb}(\eta)$
is summarized in a compact formula [eq.(\ref{rreges})]. For $L \sim
1~\mbox{Mpc}$ (today) we find $r(L,\eta) \sim 10^{-34}$ at the
Electroweak scale and $r(L,\eta)\sim 10^{-14} $  at the QCD scale,
suggesting the possibility that these primordial
seeds could be amplified by dynamo mechanisms to the values of the
magnetic fields consistent with the observed ones on these scales.

  In section II we introduce the
  model, in section III we summarize the dynamics in the different
  stages after the phase transition and discuss the dynamics of
  gauge fields including the dissipative effects of the plasma. In
  section IV we compute the spectrum of the primordial magnetic
  field generated during the different stages and assess the regime of validity of
  the approximations invoked. Our results and
  conclusions are summarized in section V.

\section{Magnetic fields in Friedmann-Robertson-Walker cosmology}

The cosmological setting in which we are primarily interested
corresponds to a symmetry breaking phase transition in a radiation
dominated Universe. Such phase transition is in principle
different from the electroweak one\footnote{If the electroweak
phase transition is weakly first order, nucleation will be almost
indistinguishable from spinodal decomposition and the phenomena
studied here may be of relevance.} and presumably occurs at a much
higher energy scale, such as the GUT scale $\sim
10^{15}\mathrm{Gev}$ but is assumed to be described by a particle
physics model that includes many fields with (hyper)-charge either
fermionic or bosonic. We will not attempt to study a particular
gauge theory phenomenologically motivated by some GUT scenario,
but will focus our study on a generic scalar field model in which
the scalar fields carry an abelian (hyper)charge. The simplest
realization of such model is  scalar electrodynamics with $N$
charged scalar fields $ \phi_r, \; r=1, \ldots, N$ and one neutral
scalar field $ \psi $ whose expectation value is the order
parameter associated with the phase transition. 

This model is inspired by the $O(4)$ linear sigma model which is the
low energy effective theory of QCD that describes chiral symmetry
breaking and the dynamics of pions\cite{dono,wilraja}. In this low
energy effective theory the {\bf neutral field} associated with $ \psi
\sim <{\bar q} q > $ acquires an expectation value while the three
pion fields $ \pi^{\pm}, \; \pi^0 $ are the (quasi)-Goldstone modes
associated with chiral symmetry breaking. The charged pions couple
minimally to the electromagnetic field \cite{dono} and obviously
chiral symmetry breaking (a nonzero expectation for the neutral field)
preserves the gauge symmetry. We argue later that the model [see
  eq.(\ref{lagra})] can describe magnetogenesis during the QCD phase
transition. The mechanisms of magnetogenesis discussed in the present
article is therefore akin to the photoproduction during the
nonequilibrium chiral phase transition \cite{prem}.

The neutral field is not coupled to the gauge field and its
acquiring an expectation value does not break the $U(1)$ gauged
symmetry. This guarantees that the abelian gauge symmetry
identified with either hypercharge  or electromagnetism is
\emph{not spontaneously broken} to describe the correct low energy
sector with unbroken $U(1)_{EM}$. We will take the neutral and the
$N$ complex (charged) fields to form a scalar multiplet under an
$O(2N+1)$ isospin symmetry. As the neutral field acquires an
expectation value this isospin symmetry is spontaneously broken to
$O(2N)$. The explicit breaking of the $O(2N+1)$ symmetry
induced by the electromagnetic coupling further reduce this symmetry
to $SU(N)\times U(1)$. If the neutral field acquires a non vanishing
expectation value, the isospin symmetry breaking does not affect the
masslessness of the photon (it will obtain a Debye screening mass from
medium effects).

The action that describes this theory in a general cosmological
background  is given by
\begin{equation}\label{lagra} S= \int d^4x
\sqrt{-g}\left[g^{\mu\nu}\left(\frac{1}{2} \partial_{\mu}\psi
 \; \partial_{\nu}\psi+ \mathcal{D}_{\mu}\phi^*  \; \mathcal{D}_\nu
\phi\right) +\mu^2 \left(\frac{\psi^2}{2}+\phi^*\phi\right)
-\frac\lambda{4N} \left(\frac{\psi^2}{2}+\phi^*\phi\right)^2
-\frac{1}{4} \mathcal{F}_{\mu \nu} \; \mathcal{F}_{\alpha \beta}
\; g^{\mu \alpha} \; g^{\nu \beta}\right]
\end{equation}
\noindent  where
\begin{equation}\label{cova}\mathcal{D}_\mu =
\partial_{\mu}-ie \mathcal{A}_\mu \quad \mbox{and} \quad
\mathcal{F}_{\mu \nu} = \partial_{\mu} \mathcal{A}_{\nu} -
\partial_{\nu}\mathcal{A}_{\mu}  \; .
\end{equation}
 and
$$
\phi^\dagger\phi=\sum_{r=1}^N\phi^\dagger_r\; \phi_r \;,\quad
D_\mu\phi^\dagger D^\mu\phi=\sum_{r=1}^N(\partial_\mu+ieA_\mu)
\phi_r^\dagger\;(\partial^\mu-ieA^\mu)\phi_r\;.
$$
Furthermore, anticipating a non-perturbative treatment of the
non-equilibrium dynamics of the scalar sector in a large $N$
expansion, we have rescaled the quartic coupling in such a way as
to display the contributions in terms of powers of $ 1/N $,
keeping $\lambda$ fixed in the large $N$ limit.

A Friedmann-Robertson-Walker line element
\begin{equation}\label{FRWds}
ds^2= dt^2-a^2(t) \; d{\vec x}^2 \; ,
\end{equation}
\noindent is conformally related to a Minkowski line element by
introducing the  conformal time $\eta$  and scale factor $C(\eta)$
as
\begin{equation}
\eta = \int \frac{dt}{a(t)}~~;~~ C(\eta) = a(t(\eta))\; ,
\end{equation}
In terms of these the line element and metric are given by
\begin{equation}\label{conformal}
ds^2 = C^2(\eta) \; (d\eta^2 - d{\vec x}^2)~~;~~g_{\mu \nu}=
C^2(\eta)  \; \eta_{\mu \nu}\; ,
\end{equation}
\noindent where $\eta_{\mu\nu}=\mbox{diag}(1,-1,-1,-1)$ is the
Minkowski metric. Introducing the conformal fields
$$
A_{\mu}(\eta,\vec x)=\mathcal{A}(t(\eta),\vec x)~;~ \Phi_r(\eta,\vec
x) = C(\eta) \; \phi_r(t(\eta),\vec x) \; , \quad 1\leq r \leq N 
~;~ \Psi(\eta,\vec x) =
C(\eta) \; \psi(t(\eta),\vec x)
$$
and in terms of the conformal time, the action now reads
\begin{equation}\label{confoS}
S= \int d\eta\; d^3x \left[\eta^{\mu
\nu}\left(\frac{1}{2}\partial_{\mu}\Psi \partial_{\nu}\Psi+
D_{\mu}\Phi^*D_{\nu}\Phi\right)-M^2(\eta)\left(\frac{\Psi^2}{2}+
\Phi^*\Phi\right)
-\frac\lambda{4N}\left(\frac{\Psi^2}{2}+\Phi^*\Phi\right)^2-\frac{1}{4}
F_{\mu \nu} \;  F_{\alpha \beta} \; \eta^{\mu \nu} \; \eta^{\alpha
\beta}\right]
\end{equation}
\noindent with
\begin{eqnarray} \label{masconf}
&&M^2(\eta) = -\mu^2 C^2(\eta)- \frac{C''(\eta)}{C(\eta)} \quad ,
\quad D_{\mu} = \partial_{\mu}-ie A_{\mu} ~~; ~~ F_{\mu \nu} =
\partial_{\mu }A_{\nu}-\partial_{\nu}A_{\mu}\; ,
\end{eqnarray}
\noindent and the primes refer to derivatives with respect to
conformal time.  Obviously the conformal rescaling of the metric
and fields turned the action into that of a charged scalar field
interacting with a gauge field in \emph{flat Minkowski
space-time}, but the scalar field acquires a time dependent mass
term\footnote{Here we neglect the effect of the conformal
anomaly\cite{dolgovanomaly}}. In particular, in the absence of
electromagnetic coupling, the equations of motion for the gauge
field $ A_{\mu} $ are those of a free field in flat space time.
This is the statement that gauge fields are \emph{conformally}
coupled to gravity and no generation of electromagnetic fields can
occur from gravitational expansion alone without coupling to other
fields or breaking the conformal invariance of the gauge sector.
The generation of electromagnetic fields must arise from a
coupling to other fields that are not conformally coupled to
gravity, or by adding extra terms in the Lagrangian that would
break the conformal invariance of the gauge
fields\cite{turnerwidrow}.

The conformal electromagnetic fields
$\vec{\mathcal{E}},\vec{\mathcal{B}}$ are related to the physical 
$\vec E, \vec B$ fields by the conformal rescaling
\begin{equation}\label{physicalfields} \vec{E}=
\frac{\vec{\mathcal{E}}}{C^2(\eta)}~~;~~\vec{B}=
\frac{\vec{\mathcal{B}}}{C^2(\eta)} \; ,
\end{equation}
\noindent corresponding to fields of scaling dimension two. A
gauge invariant formulation leads to the following Lagrangian
density (for details see\cite{magfiI,gauginv})
\begin{eqnarray}\label{gauginvlag} {\cal L}= && \frac{1}{2}\partial_{\mu}\Psi\;
\partial^{\mu}\Psi+\partial_{\mu}\Phi^\dagger   \;
\partial^{\mu}\Phi+\frac{1}{2}\partial_{\mu}\vec A_T\cdot\partial^{\mu} \vec
A_T+\frac{1}{2} (\nabla A_0)^2
-M^2(\eta)\left(\frac{1}{2}\Psi^2+\Phi^\dagger\Phi\right)
-\frac\lambda{4N}
\left(\frac{1}{2}\Psi^2+\Phi^\dagger\Phi\right)^2 \nonumber \\&&
-ie\vec A_T\cdot\left(\Phi^\dagger\nabla\Phi-
\nabla\Phi^\dagger\Phi\right)- e^2(\vec
A_T^2-A_0^2)\;\Phi^\dagger\Phi-ie\; A_0\left(
{\dot\Phi}^\dagger\Phi-\Phi^\dagger\dot\Phi\right) \; ,
\end{eqnarray}
\noindent where $\Phi$ is a gauge invariant \emph{local} field
which is non-locally related to the original fields, and
$\vec{A}_T$ is the transverse component of the vector field
(${\vec \nabla}\cdot {\vec A}_T=0$) and  $A_0$ is a
non-propagating field as befits a Lagrange multiplier, its
dynamics is  completely determined by that of the  charge
density~\cite{magfiI,gauginv}.

The main point of this discussion is that the framework to obtain
the power spectrum of the generated magnetic field presented below
is fully \emph{gauge invariant}.

The theory described by the Lagrangian eq.(\ref{confoS}) above bears
some similarity to the scalar-gauge field theory that describes
semilocal strings\cite{vachaspati,achucarro,hindmarsh}, however
there are important differences between the model studied here and
that studied in refs.\cite{vachaspati,achucarro,hindmarsh}: i) we
assume that symmetry breaking occurs along the neutral direction
thus the charged scalar field \emph{does not} acquire an
expectation value, whereas in the semilocal theory of
refs.\cite{vachaspati,achucarro,hindmarsh} the charged fields
acquire an expectation value, and the gauge symmetry is
spontaneously broken. ii) We study the dynamics beginning from an
initial state in LTE above the critical temperature, follow the
dynamics \emph{through} the phase transition and compute
systematically to lowest order in $\alpha_{em}$ the
non-equilibrium spectrum of magnetic fields generated by the
process of phase separation. The goal of the studies in
ref.\cite{achucarro} are very different focusing on the rate of
production of semilocal strings. The initial state studied in
these references places the scalar field at the minimum of the
(classical) potential and the phases are distributed at random,
with particular initial conditions on the gauge fields, namely
conditions corresponding to zero temperature, broken symmetry
states. Furthermore the dynamics in ref.\cite{achucarro} is
studied in flat space time with an \emph{ad hoc} dissipative term
for the scalar field.

In contrast, we study the full quantum dynamics beginning from a
state of LTE above $T_c$ evolving the quantum Heisenberg equations
of motion and calculate the magnetic field consistently to lowest
order in $\alpha$. Thus while the theory studied here and that
proposed in refs.\cite{vachaspati,achucarro,hindmarsh} bear a
resemblance, they describe very different physics and we study a
different set of phenomena.

\section{Phase transitions in radiation dominated cosmology}

\subsection{Kinematics}

We consider a phase transition corresponding to the dynamics of
small field models where the scalar field has vanishing
expectation value but with a symmetry breaking potential, namely
at the top of the potential hill.

In a radiation dominated cosmology, the initial state is that of
local thermodynamic equilibrium at an initial temperature $T
>>T_c$. Using finite temperature field theory in
an expanding background geometry, it is shown~\cite{scaling} that
the effective time dependent mass term depends on the effective
time dependent temperature $T(t)=T/a(t)$ which reflects the
cooling from the cosmological expansion (see below). Hence at a
given time the temperature equals the critical and the phase
transition occurs. Field modes with wavectors much larger than the
symmetry breaking scale $\mu$ will remain in LTE and will not be
affected by the symmetry breaking dynamics\cite{boysinglee}.

We normalize the scaling factor $C(\eta)$ at the reheating time
$\eta=\eta_R$ in such a way that $C(\eta_R)=1$ then the explicit
expression for $C(\eta)$ reads
\begin{equation}\label{C-rde}
C(\eta)=H_R\;\eta
\end{equation}
\noindent where $H_R$ is the Hubble constant at the reheating
time, $H_R=\eta_R^{-1}$.

We can relate $H_R$ to the reheating temperature and the Planck
mass $ G^{-\frac12} $ through the  equation
\begin{equation}\label{blackbody}
\rho= \frac{\pi^2g_*}{30}~T_R^4
\end{equation}
and the Einstein-Friedman  equation
\begin{equation}\label{H-T}
H_R=\left(\frac83\pi \; G \;
\rho\right)^{1/2}=\frac{T_R^2}{M_*} \; ,
\end{equation} \noindent
where $g^*$ is the effective number of degrees of freedom at the
reheating temperature and we introduced the scale $M_*$ of the
order of the Planck mass
\begin{equation}\label{Mstar}
M_*=\frac{3 \; \sqrt{5}}{2 \; \pi^\frac{3}{2}}\;
\frac{1}{\sqrt{g_* \; G}} \; .
\end{equation}
In radiation dominated epoch the time-dependence of the mass term
(\ref{masconf}) is given by the expression
\begin{equation}\label{mass}
-M^2(\eta)=\mu^2 \; H_R^2 \; \eta^2=\tilde\mu^4 \; \eta^2 \; ,
\end{equation}
\noindent where we see the emergence of a new mass scale
\begin{equation}\label{mu-tilde}
\tilde\mu=\sqrt{\mu \;  H_R}\;.
\end{equation}
This scale will play an important role in the following discussion
and in the comparison with results obtained in  Minkowski
space-time\cite{magfiI}. There is a last scale which plays a
relevant role, the horizon scale $r_H(\eta)$ which is fixed by the
evolution on the time of the Hubble constant:
\begin{equation}\label{hubble}
r_H(\eta)=\frac{1}{H(\eta)}=C(\eta)~\eta=H_R~\eta^2 \; .
\end{equation}
Modes with physical wavelength
$\lambda_{phys}=\frac{2\pi}{k_{phys}}$ inside the horizon
\begin{equation}
\lambda_{phys}(\eta)\sim k_{phys}^{-1}(\eta)<r_H(\eta)
\end{equation}
are causally connected, modes outside the horizon are causally
disconnected.

The relaxation rate of hard modes of the charged fields is given
by\cite{scalarqed}
\begin{equation}\label{relrate}
\Gamma(\eta) \sim \alpha \; T(\eta) \; \ln\frac{1}{\alpha} \; ,
 \end{equation}
where the effective temperature varies with time as
\begin{equation}\label{Tef}
 T(\eta)=\frac{T_R}{C(\eta)}
\end{equation}
and  the expansion rate given by
\begin{equation}
H(\eta)=\frac{T^2(\eta)}{M_{*}} \; .
\end{equation}
Therefore,
\begin{equation} \frac{\Gamma(\eta)}{H(\eta)} \sim
\frac{10^{16}}{T(\eta)[\mbox{Gev}]} \; .
\end{equation}
Thus hard modes are in thermal equilibrium for $T(\eta)\leq
10^{15}\mbox{Gev}$.

In particular, modes with $k\sim T_R$ are the hard modes that give
the leading contribution to the conductivity in the high
temperture limit\cite{baym,yaffe}. Modes with $k < \mu$ will
manifest the long-wavelength spinodal instabilities and their
dynamics will be strongly out of
equilibrium\cite{boysinglee,nuesfrw,nuestros,destri}. Their
amplitude becomes non-perturbatively
large\cite{boysinglee,nuesfrw,nuestros,destri} and will be
responsible for the non-equilibrium generation of the primordial
magnetic field\cite{magfiI}.

Using eqs.(\ref{C-rde}) and (\ref{Tef}) we can write the conformal
time as
\begin{equation}\label{eta.T}
\eta = \frac{T_R}{H_R \; T(\eta)} \; .
\end{equation}
As it will become clear below an important cosmological  quantity
is the product
\begin{equation}\label{k.phys}
k\;\eta=\frac{k}{C(\eta)} \; \mathcal{C}(\eta) \;
\eta=\frac{k_{phys}(\eta)}{T(\eta)}\;T(\eta) \;
r_H(\eta)=\frac{2\pi}{LT_R}  \;   \frac{M_*}{T(\eta)} \; .
\end{equation}
The ratio
\begin{equation}
\frac{k_{phys}(\eta)}{T(\eta)}=\frac{2\pi}{L \; T_R}
\end{equation}
is a kinematical invariant. Its value today is determined by the
scale $L$ which will be typically chosen to correspond to a
galactic scale or the scale of galaxy clusters, and the
temperature of the CMB. It is given by,
\begin{equation} \label{LTR}
L \; T_R = 3.7\times 10^{25}\left(\frac{L}{\mbox{Mpc}}\right) \; .
\end{equation}
Therefore,
\begin{equation} \label{horiz}
k \; \eta \sim 10^{-9} \; \frac{T_{EW}}{T(\eta)}\left(\frac{\mbox{Mpc}}{L}
\right)=\left\{
\begin{array}{l}
  10^{-22}~ \; \mbox{for}~T(\eta)=T_R \sim 10^{15} \; \mbox{Gev} \\
  10^{-9}~ \; \mbox{for\, the\, EW\, transition} \\
  10^{-6}~ \; \mbox{for\, the\, QCD\, transition}
\end{array}\right.
\end{equation}
for $L\sim 1~\mbox{Mpc}$. Thus, during the regime of interest in
this article, $k\eta \ll 1$ for scales of galaxy clusters. A
noteworthy aspect of eq.(\ref{horiz}) is that the wavelengths
corresponding to the scale of galaxies or clusters today were well
outside the horizon during the radiation dominated era when the
electroweak and QCD phase transitions occurred.

Another important quantity is the ratio of the wavevector $k$ of
the primordial magnetic field to the conductivity.

As it will be discussed below, the physical conductivity is given
by
\begin{equation}\label{conduc}
\sigma(\eta) = \frac{\mathcal{C}  \;  N(\eta) \; T(\eta)}{\alpha
\;  \ln\frac{1}{\alpha N(\eta)}} \; ,
\end{equation}
\noindent where $\mathcal{C}$ is a constant of $\mathcal{O}(1)$,
 $N(\eta)$ is the number of ultrarelativistic charged species,
and we have neglected the (logarithmic) dependence on the energy
scale in the running coupling constant. For this discussion we
will neglect  the time dependence  of $N(\eta)$  assuming that the
number of charged ultrarelativistic species remains constant (this
assumption can be relaxed without qualitative modifications of the
main argument). Under this assumption
\begin{equation}\label{physisigma}
\sigma(\eta)= \frac{\sigma_R}{C(\eta)} \; ,
\end{equation}
\noindent with $\sigma_R$ being the \emph{comoving} conductivity
determined at the time of reheating
\begin{equation}\label{comovingsigma}
\sigma_R= \frac{\mathcal{C} \;  N \;  T_R}{\alpha \ln\frac{1}{\alpha N}} \; .
\end{equation}
Thus the ratio,
\begin{equation}\label{otroratio}
\frac{k_{phys}(\eta)}{\sigma(\eta)}\sim
\frac{2\pi\alpha}{N\;LT_R}\sim 10^{-27} \left(\frac{\mbox{Mpc}}{L}
\right)\; ,
\end{equation}
\noindent neglecting logarithmic corrections.

Furthermore,
\begin{equation}\label{sigrat}
\sigma(\eta) \;  r_H(\eta) \sim \frac{M_*}{\alpha T(\eta)} \sim
\left\{ \begin{array}{l}
  10^{5} \; ~\mbox{for}~T(\eta)=T_R \sim 10^{15}\mbox{Gev} \\
  10^{18} \; ~\mbox{for\,the\,EW\,phase\,transition} \\
  10^{21} \; ~\mbox{for\,the\,QCD\,phase\,transition}
\end{array}\right.
 \end{equation}
\noindent where we have neglected logarithmic corrections.
Therefore $\sigma_R \; \eta \gg 1$ throughout the radiation
dominated era considered in this article. The regime $\sigma_R  \;
\eta \gg 1~;~~ k^2 \; \eta/\sigma_R \ll 1$ is dominated by the
(slow) hydrodynamic relaxation of the magnetic field.

Another relevant estimate involves the (comoving) \emph{diffusion
length} $\xi_{diff}(\eta)=\sqrt{\eta/\sigma_R}$
\begin{equation}\label{difflength}
\frac{\xi_{diff}(\eta)}{\eta} \sim \sqrt{\frac{\alpha \;
T(\eta)}{M_*}}\sim\left\{\begin{array}{l}
  10^{-3} \; ~\mbox{for}~T(\eta)=T_R \sim 10^{15}\mbox{Gev} \\
  10^{-9} \; ~\mbox{for\,the\,EW\,phase\,transition} \\
  10^{-10} \; ~\mbox{for\,the\,QCD\,phase\,transition}
\end{array}\right.
\end{equation}
\noindent where again we have neglected logarithmic terms.
Therefore the diffusion length is much smaller than the Hubble
radius during the radiation dominated era. Finally, combining
eqs.(\ref{difflength}) and (\ref{horiz}) we find
\begin{equation}\label{produ}
10^{-25}\leq k \; \xi_{diff}(\eta) \leq 10^{-16}\; ,
\end{equation}
\noindent between reheating and the time of the QCD phase
transition.

The contribution from the hard modes of both the charged scalar
and gauge fields which remain in local thermodynamic equilibrium
lead to an effective mass for the scalar field. This
\emph{thermal} mass is obtained from the long-wavelength limit of
the scalar field self-energy and includes the hard thermal loop
contributions from the gauge and scalar
fields\cite{kapusta,lebellac}. This thermal mass is given by
\begin{equation}\label{thermalmass}
m^2_{T} = \frac{T^2_R}{24} \left(\lambda  + 3 \; e^2\right) \; .
\end{equation}
 Finally, another important quantity is the Debye
screening length that determines the scale at which long-range
forces are screened by the polarizability of the medium. In an
ultrarelativistic plasma, the comoving Debye screening length is
given by\cite{kapusta,lebellac}
\begin{equation}\label{debyelength}
\xi_D \sim \frac{1}{e \; T_R}
\end{equation}
the ratio of the Debye screening length to the Hubble radius is
given by
\begin{equation}
\frac{\xi_D}{d_H} \sim \frac{1}{e} \;  \frac{T(\eta)}{M_*}
\end{equation}
Hence $\xi_D \ll d_H(\eta)$ for $T(\eta) \leq 10^{16}\mbox{Gev}$,
thus long range forces are screened over very short distances. The
formation of long-wavelength domains with typical size of the
order of the Hubble radius\cite{scaling} leading to strong charge
and current fluctuations that will seed magnetic fields, will not
be hindered by long-range forces, which are effectively screened
over sub-horizon distances.

Magnetic field generation via charge asymmetries during a period
in which electromagnetism was spontaneously broken was previously
studied by Dolgov and Silk\cite{dolgovsilk} who argued that
long-range forces would be screened by the Higgs mechanism.
This is different from the situation studied in this article,
where the $U(1)$ symmetry associated with electromagnetism (rather
hypercharge) is \emph{not} spontaneously broken. Long range forces
are screened by the plasma, a situation not considered
in\cite{dolgovsilk}.

\subsection{Scalar fields dynamics}

For completeness and to highlight the  aspects of the
non-equilibrium dynamics most relevant to the generation of
magnetic fields, we summarize the main features of scalar field
dynamics. For further details the reader is referred
to~\cite{nuesfrw,nuestros,destri}. In what follows we will neglect
the backreaction of the gauge fields on the dynamics of the scalar
fields. The rationale for this is that the main non-equilibrium
processes that lead to magnetogenesis will be non-perturbative in
the \emph{scalar} sector and result from the instabilities
associated with the phase transition. The contribution from the
gauge fields, in the form of self-energies for the scalar fields,
do not feature the instabilities associated with the phase
transition and will, furthermore, be suppressed at least by one
power of $\alpha$ the (hyper) electromagnetic coupling constant as
compared to the scalar self-interaction.

As described above, the non-equilibrium evolution of
long-wavelength modes begins with the  spinodal instabilities
which  result in an exponential growth of the amplitudes for
long-wavelength fluctuations. When the non-linearity becomes of
the same order as the tree-level terms in the equations of motion,
the back reaction of these fluctuations  shuts-off the
instabilities~ \cite{nuesfrw,nuestros,destri}.  Therefore a
non-perturbative treatment of the dynamics is required. The large
$N$ limit of the scalar sector allows a systematic
non-perturbative treatment of the dynamics which is renormalizable
and  maintains the conservation laws\cite{nuesfrw,nuestros}.

We will therefore study the dynamics in leading order in the large
$N$ limit that already reveals the important non-equilibrium
features of the evolution.

{\bf Radiative corrections: }

The contribution from the gauge fields to the equations of motion
of the long-wavelength modes of the scalar fields arise through
self-energy corrections. To lowest order in $\alpha$ these are
dominated by the hard modes of the gauge fields with momenta $\sim
T$ (hard thermal loops) which lead to a contribution to the
thermal mass given by $eT/\sqrt{8}$~\cite{kapusta,lebellac}. Thus
the lowest order radiative corrections  had already been accounted
for in the thermal mass eq.(\ref{thermalmass}).

The \emph{non-equilibrium} effects in the gauge contribution of
the scalar self-energy will arise from polarization loops in the
photon propagator. This in turn will induce non-equilibrium
radiative corrections to the self-energy of the scalar fields 
of the order $\mathcal{O}(\alpha^2)$. These small contributions can
be safely neglected in this context. Thus, to this order the radiative
corrections to the scalar field from the gauge field propagator in the scalar
self-energy  are accounted for in the thermal mass.

Hence the dynamics
of the scalar field is studied along the same lines as presented
in refs.~\cite{destri,nuesfrw,nuestros} but the only difference is
in the initial conditions in the modes that reflect the thermal
mass in LTE.

Since symmetry breaking is chosen along the direction of the
neutral field $\Psi$ we write
\begin{equation} \Psi(\vec x,\eta ) =
\sqrt{N} \; \varphi(\eta ) + \chi(\vec x,\eta ) \quad ; \quad \langle
\chi(\vec x,\eta ) \rangle = 0 \label{expecval}
\end{equation}
\noindent where the expectation value is taken in the time evolved
density matrix or initial state. The leading order in the large $N$
limit is obtained either by introducing an auxiliary field and
establishing the saddle point or equivalently by the
factorizations\cite{nuesfrw,nuestros}
\begin{eqnarray}
&&(\Phi^{\dagger}\Phi)^2 \rightarrow 2 \langle
\Phi^{\dagger}\Phi\rangle \Phi^{\dagger}\Phi \cr \cr
&&\chi \Phi^{\dagger}\Phi \rightarrow \chi \langle
\Phi^{\dagger}\Phi\rangle \nonumber \; .
\end{eqnarray}
This factorization that yields the leading
contribution in the large $ N $ limit makes the Lagrangian for the
scalar fields quadratic (in the absence of the gauge coupling) at
the expense of a self-consistent condition: thus charged fields
$\phi$ acquire a self-consistent time dependent mass.

The dynamics is determined by the Heisenberg equations of motion of the neutral
field $ \Psi$ and the charged fields $ \Phi
$~\cite{nuesfrw,nuestros,destri}.  We will consider that at
the onset of the radiation dominated era, the system is in the
symmetric high temperature phase in local thermal equilibrium with
a vanishing expectation value for the scalar fields. Consequently, the
initial conditions are $ <\Psi(\vec x, 0)> = \sqrt{N} \; \varphi(0) =
0 , \; <{\dot \Psi}(\vec x, 0)> = \sqrt{N} \; {\dot\varphi}(0) = 0 ,
\; <\Phi_r(0,\vec x)> = 0, \;  <{\dot \Phi}_r(0,\vec x)> = 0 $.

In the absence of explicit symmetry breaking perturbations the
expectation value of the scalar field will remain zero throughout
the evolution, thus $\varphi\equiv 0$.

It is convenient to introduce the mode expansion of the charged
fields
\begin{equation}
\Phi_r(\eta ,\vec x)= \int \frac{d^3 k}{\sqrt{2 \, (2\pi)^3}}
 \left[ a_r(\vec k) \; f_k(\eta )\;
e^{i\vec k\cdot \vec x}+ b_r^\dagger(\vec k)\; f^*_k(\eta )\;
e^{-i\vec k\cdot \vec x} \right]\quad  , \quad r=1,\ldots,N \; ,
\label{phidecompo}
\end{equation}
with $ < a_r(\vec k) > =  <b_r(\vec k)> = 0 $.

In leading order in the large $N$ limit, the Heisenberg equations
of motion for the charged fields translate into the following
equations of motion for the mode functions and the expectation value
of the neutral field for $\eta>\eta_R$,~\cite{nuesfrw,nuestros,destri}
 \begin{eqnarray}
&&\left[\frac{d^2}{d\eta ^2}+k^2-M^2(\eta)+\frac{\lambda}{2}\varphi^2(\eta )+
\frac{\lambda}{2N}\langle\Phi^{\dagger}\Phi\rangle\right ] \;
f_k(\eta )=0 \; , \cr \cr
&&\left[\frac{d^2}{d\eta ^2}-M^2(\eta)+\frac{\lambda}{2}\varphi^2(\eta )+
\frac{\lambda}{2N}\langle\Phi^{\dagger}\Phi\rangle\right ] \;
\varphi(\eta )=0 \;.  \label{unsaledeqnsofmot}
\end{eqnarray}
Obviously, the initial conditions $ \varphi(0) = 0 , \;
{\dot\varphi}(0) = 0 $ imply that $ \varphi(\eta ) = 0 $ for all times.
That is, $ \varphi(\eta ) = 0 $ is a fixed point of the dynamics.

We must now append initial conditions for  the mode functions
$f_k(\eta )$.  The initial conditions on the mode functions
$f_k(\eta )$ depend on the value of the wavevector $k$ as compared
to the horizon scale $H_R^{-1}$ at the reheating time:
\begin{itemize}
\item{$\mathbf{k>H_R}$:} for fluctuations inside the horizon, we
may assume thermal quasi-particle boundary conditions at the
reheating temperature:
\begin{equation}
f_k(\eta_R)= \frac{1}{\sqrt{W_k}} \quad ; \quad{f'}_k(\eta_R)=
-iW_k~f_k(\eta_R)\; , \quad W_k=\sqrt{k^2+m^2_T}\; ,
\label{iniconds}
\end{equation}
\noindent where the frequencies $W_k$ are quasi-particle
frequencies with thermal mass $m^2_T$ given by eq.(\ref{thermalmass})
at the reheating temperature,
$$
W_k=\sqrt{k^2+m^2_T} \; .
$$
For these modes the assumption of local thermodynamic equilibrium
is well motivated and we have
\begin{equation}
\langle a_r^\dagger(\vec k) \; a_s(\vec k)\rangle =\langle
b^\dagger_r(k) \; b_s(k)\rangle =\delta_{rs} \; n_k~~;~~
n_k=\frac1{e^{\frac{W_k}{T}}-1}\label{BE.scalars}
\end{equation}
\item{$\mathbf{k<H_R}$:} for superhorizon fluctuations, which are
causally disconnected at the reheating time, we cannot assume a
thermalized distribution. The correct distribution has to  be
derived by following the dynamics from the inflationary stage,
when the fluctuations were well inside the horizon. While a
complete discussion of the initial conditions is left to a
forthcoming article, the case under consideration we will see that
the dependence on the initial conditions is rather weak and only
during the initial stages of the phase transition. For the later
stages, dominated by the scaling solution described below, the
dynamics is \emph{universal} and does not depend on the initial
conditions. We will simply assume that both $W_k$ and $n_k$ have a
finite non-zero limit as $k\to0$ namely the only important
quantities  for the dynamics of long-wavelength fluctuations are
\begin{equation}\label{k->0 bc} \lim_{k\to 0}W_k= W_0\;,\quad
0<W_0<\infty,\quad \lim_{k\to 0}n_k= n_0\;,\quad 0<n_0<\infty\;.
\end{equation}
\end{itemize}

With this choice of the initial state we find the backreaction
term to be given by
\begin{equation}\label{backreaction}
\frac{\lambda}{2N}\langle\Phi^{\dagger} \Phi\rangle =
\frac{\lambda}{4}\int \frac{d^3k}{(2\pi)^3}\;|f_k(\eta
)|^2[1+2n_k]\;.
\end{equation}
This expectation value is ultraviolet divergent, it features
quadratic and logarithmic divergences in terms of an upper
momentum cutoff. The quadratic divergence and part of the logarithmic
divergence (the one proportional to the mass term) are absorbed in
a renormalization of the mass term $\mu^2 \rightarrow \mu^2_R$ and
the remainder logarithmic divergence is absorbed into a
renormalization of the scalar coupling $\lambda \rightarrow
\lambda_R$. While these aspects are not relevant for the
discussion here, they are mentioned for completeness, the reader
is referred to \cite{scaling} for details.

After renormalization the self-consistent field
$\frac{\lambda}{2N}\langle \Phi^\dagger \Phi\rangle$ is subtracted
twice, and is given by (for details see~\cite{scaling} and
references therein)
\begin{eqnarray}\label{JplusI}
\frac{\lambda}{2N}\langle \Phi^\dagger \Phi\rangle &=&
\lambda_R  \; \left[J(\eta)+I(\eta)\right] \quad , \quad
\lambda_R \; J(\eta) = \frac{\lambda_R}{4\pi^2}\int_0^{\infty}
q^2\;dq \;  {|f_q(\eta )|^2}~n_q \; , \cr \cr
\lambda_R  \; I(\eta)&=&\frac{\lambda_R}{8\pi^2}\int_0^{\infty}
q^2\;dq \; \Biggr\{{|f_q(\eta)|^2}-\frac{1}{q}
+\frac{\Theta(q-K^2)}{2q^3}\biggl[-\mu^2_R+
\frac{\lambda}{2N}\langle \Phi^\dagger \Phi\rangle\biggr]
        \Biggr\}. \label{gsigma}
\end{eqnarray}
\noindent and the mass and coupling are replaced by their
renormalized counterparts $\mu^2_R;\lambda_R$ respectively. Here
$K$ is an arbitrary renormalization scale. In order to avoid
cluttering of notation we now drop the subscript $R$ for
renormalized quantities, in what follows $\mu;\lambda$ stand for
the renormalized quantities.

The finite temperature term $J(\eta)$ has contributions from short
wavelengths for which the mode functions are of the form
$f_q(\eta) \sim e^{iq\eta}/\sqrt{q}$ and contributions from long
wavelengths. The contribution from short wavelengths is the same
as that in equilibrium in  Minkowski space time and determines the
hard-thermal loop~\cite{kapusta,lebellac} contribution to the
self-energy given by\cite{scaling}
\begin{equation}\label{Jhtl}
J_{HTL}= \frac{T^2_R}{24}
\end{equation}
\noindent where we have used that the short wavelength modes are
in thermal equilibrium at the reheating temperature $T_R$. This
hard thermal loop contribution has been self-consistently
accounted for in the thermal mass of the scalar field eq.(\ref{thermalmass}).

It is convenient to separate the hard thermal loop component
eq.(\ref{Jhtl}) from eq.(\ref{JplusI}) and define
\begin{equation}\label{sigma}
\lambda\Sigma(\eta)= \frac{\lambda}{8\pi^2}\int_0^{\infty} q^2\;dq
\Biggr\{{|f_q(\eta)|^2}(1+2n_q)-\frac{1}{q}\left[1+
\frac{2}{e^{\frac{q}{T_R}}-1}\right]
+\frac{\Theta(q-K^2)}{2q^3}\biggl[-\mu^2+ \lambda\Sigma(\eta
)\biggr] \Biggr\}.
\end{equation}
After renormalization and in terms of dimensionless quantities,
the non-equilibrium dynamics of the charged scalar fields is
completely determined by the  following equations of
motion~\cite{scaling,nuesfrw,nuestros,destri},
\begin{eqnarray}\label{ecmov}
&&\left[\frac{d^2}{d\eta ^2}+
\mathcal{M}^2(\eta)+q^2+\lambda\Sigma(\eta ) \right]f_q(\eta
)=0~~;~~ f_q(\eta_R)=\frac{1}{\sqrt{W_q}}~~;~~{f'}_q(\eta_R)
=-iW_q~f_q(\eta_R)  \label{modeeqnew}
\end{eqnarray}
\noindent with $W_q$ given by eq. (\ref{mass}) and the effective,
(conformal) time dependent mass is given by
\begin{eqnarray}\mathcal{M}^2(\eta)&=&
C^2(\eta) \; \mu^2\left[\frac{T^2_R}{C^2(\eta) \;
T^2_c}-1\right]\label{mossoft}\\
T^2_c &=& \frac{24 \; \mu^2}{\lambda+3e^2} .\label{crittemp}
\end{eqnarray}
Where we have used $ \varphi = 0 $ as a fixed point of the dynamics.

The time dependent mass term $\mathcal{M}^2(\eta)$ includes the
high temperature corrections and clearly displays the cooling
associated with the expansion in the form of a time dependent
effective temperature $T_{eff}(\eta)=T_R/C(\eta)$.  The phase
transition occurs at a time $\eta_c$ when
$T_{eff}(\eta_c)=T_c$, thus for $\eta > \eta_c$ the
effective time dependent mass term is $\mathcal{M}^2(\eta) =
M^2(\eta)= -\mu^2 \; H^2_R  \; \eta^2=-\tilde{\mu}^4  \; \eta^2$
as given by equations (\ref{mass})-(\ref{mu-tilde}).

The full time evolution of mode functions in a radiation dominated
cosmology has been studied analytically and numerically in detail
in ref.\cite{nuesfrw,scaling}. Here we highlight the most
important features which are necessary ingredients to study
magnetogenesis. The reader is referred to\cite{scaling} for a more
comprehensive discussion.

The are two main dynamical stages in the evolution:

\begin{itemize}
\item{Spinodal stage: this is the stage immediately after the
phase transition which is dominated by spinodal decomposition and
the growth of long-wavelength fluctuations\cite{magfiI}.  This
stage spans the time scale $\eta_c\leq \eta \leq \eta_{nl}$
where the non-linear time scale $\eta_{nl}$ is determined by (see
below)
\begin{equation}\label{etanl} \eta^2_{nl} =
\frac{\lambda\Sigma(\eta_{nl})}{\tilde{\mu}^4 }
\end{equation}
During this stage the back-reaction, determined by the term
$\lambda \; \Sigma(\eta)$, can be neglected and the dynamics is
\emph{linear}. }

\item{Scaling stage: This is a stage in which the non-linearity
encoded by the back-reaction term $\lambda\Sigma(\eta)$ are very
important and compete with the tree level term in the equations of
motion. This stage is described by a scaling solution of the
equations of motion for the modes with small wavevectors and
describes the non-equilibrium relaxation of long-wavelength
fluctuations\cite{scaling,turok,durrer}. }

\end{itemize}

\subsubsection{Spinodal stage}\label{sec:spino}

After the phase transition but before the non-linear time scale
after which the back-reaction becomes important, namely for
$\eta_c \ll\eta<\eta_{nl}$ the time dependent mass
term is given by $\mathcal{M}^2=-\tilde{\mu}^4 \eta^2$, and for
weak coupling $\lambda \ll 1$ we can neglect the back-reaction
$\lambda\Sigma(\eta )$. The equations of motion for the mode
functions during this stage are given by
\begin{equation}\label{parab.cylinder}
\left[\frac{d^2}{d\eta ^2}+q^2- \tilde\mu^4\eta^2\right]f_q(\eta
)=0\quad ;\quad q<\tilde\mu^2\eta
\end{equation}
We note that for $T_c\sim 10^{15} \mbox{GeV}$
\begin{equation}\label{muetac}
\tilde{\mu} \; \eta_c =\left(\sqrt{\frac{\lambda+3\;e^2}{24}}
\frac{M_*}{T_c}\right)^{1/2}\sim 10
\end{equation}
and therefore for $\eta > \eta_c$ we are in the regime
$\tilde{\mu} \; \eta  \gg 1 \; .$
It is clear  that the mode functions $f_q(\eta)$ will increase
exponentially in the band of \emph{unstable} wavevectors
$q<\tilde{\mu}^2\eta$. Eq.(\ref{parab.cylinder}) can be solved
exactly in terms of Hermite functions\cite{gr}
\begin{equation}\label{hermi}
f_q(\eta )= b_q \; e^{- \frac12( \tilde{\mu} \; \eta)^2} \;
H_{\frac12\left( \frac{q^2}{\tilde{\mu}^2}-1\right)}(\tilde{\mu}
\; \eta) + a_q \; e^{\frac12( \tilde{\mu} \; \eta)^2} \;
H_{-\frac12\left( \frac{q^2}{\tilde{\mu}^2}+1\right)}(i \,
\tilde{\mu} \; \eta)
\end{equation}
where the constants $ a_q $ and $ b_q $ are fixed by the initial
conditions (\ref{ecmov}). For $ \tilde{\mu} \; \eta \gg 1 $ we can
use the asymptotic behavior of the Hermite functions\cite{gr},
$$
H_{\nu}(z) \buildrel{z \gg 1}\over = (2\, z)^{\nu}\left[ 1 + {\cal
O}\left(\frac{1}{z^2}\right) \right]
$$
and we find for the mode functions,
\begin{equation}\label{asimod}
f_q(\eta )\buildrel{\tilde{\mu} \; \eta \gg 1}\over = a_q \;
e^{\frac12( \tilde{\mu} \; \eta)^2} \; \left(\tilde{\mu} \;
\eta\right)^{ -\frac{q^2}{\tilde{\mu}^2}-1} \left[ 1 + {\cal
O}\left(\frac{1}{ \tilde{\mu}^2 \; \eta^2}\right] \right]
\end{equation}
Since the exponentially damped solution becomes negligible the
phases of the mode functions $f_q(\eta )$ {\em freeze}, namely,
they become constant in time and are  slowly varying functions of
$q$ for long wavelengths.

This is very similar to the situation in Minkowski space-time,
where the mode functions however increase as $e^{\mu t}$, i.e.
much slower. In any case the soft ($q\to 0$) modes are the most
amplified at the end of the evolution, therefore, the quantum
fluctuations (\ref{sigma}) are dominated by the lower integration
bound $ q = 0 $.

We notice that the freezing of the long-wavelength mode functions
will play an important role in the discussion about the magnetic
field generation, since it assures the independence of the final
result from the initial particle distribution function, except for
subleading corrections.

The physics of the phase transition is essentially the same as in
Minkowski space-time\cite{magfiI,nuesfrw,nuestros,destri}, since
the exponential growth of modes in the spinodally unstable band
will make the back reaction term $ \lambda \Sigma(\eta) $ begin to grow and
eventually cancel the term $-\tilde{\mu}^4\eta^2$ in the equations
of motion (for $\eta >> \eta_c$ the effective time dependent
temperature vanishes).

This will happen at a \emph{non-linear} time scale defined
by~\cite{nuesfrw,nuestros}
\begin{equation}\label{spin}
\lambda \; \Sigma(\eta_{nl}) = \tilde\mu^4 \; \eta_{nl}^2
\end{equation}
Two important aspects are described by $\eta _{nl}$: i) at this
time scale the phase transition is almost complete since $\lambda
\Sigma(\eta_{nl}) = \tilde\mu^4 \; \eta_{nl}^2$ means that
$\lambda \langle \Phi^\dagger \Phi \rangle /2N =
\tilde\mu^4\eta_{nl}^2$, namely the mean square root fluctuations
in the scalar field probe the manifold of minima of the potential.

 ii) At $\eta  \sim \eta _{nl}$  the mean square root fluctuations
  of the field are of order $M^2(\eta_{nl})
/{\lambda}$ probing the vacuum manifold,  and the non-linearities
become very important. The back reaction $\lambda \Sigma(\eta_{nl})$
becomes comparable to $M^2(\eta_{nl})$ and the instabilities
shut-off. Thus for $\eta_c< \eta  < \eta _{nl}$ the dynamics is
described by the \emph{linear} spinodal instabilities while for
$\eta>\eta_{nl}$ a full non-linear treatment of the evolution is
required. As it will be discussed below this later stage is
described by the emergence of a scaling solution.

For $\eta_{nl} > \eta \gg \tilde\mu^{-1}$ the asymptotic form
(\ref{asimod}) for the mode functions apply and we find for the
the quantum fluctuations (\ref{sigma}) which dominated by the
lower integration bound $ q = 0 $,
\begin{equation}\label{sigts}
\lambda \; \Sigma(\eta_{nl}) = \lambda \; (1+2\,n_0) \;
\frac{{\tilde\mu}^2 \; |a_0|^2}{32 \; \pi^{\frac52}} \;
\frac{e^{\tilde\mu^2\eta_{nl}^2} }{\tilde\mu \; \eta_{nl} \;
\left[\ln\left({\tilde\mu}\;\eta_{nl}\right)\right]^{\frac32}
}\left[ 1 + {\cal O}\left(\frac{1}{{\tilde\mu}\;\eta_{nl}}
\right)\right] \; .
\end{equation}
\noindent This leads to the following estimate for the spinodal
time for weak coupling~$\lambda$
\begin{equation}\label{spinotime}
\eta^2_{nl} = \frac{1}{\tilde\mu^2}\left[ \ln\left(\frac{32 \;
\pi^{\frac52}}{\lambda \; (1+2\,n_0)\;|a_0|^2\;\tilde\mu}\right) +
\frac32 \; \ln \ln\left(\frac{32 \; \pi^{\frac52}}{\lambda \;
(1+2\,n_0)\;|a_0|^2\;\tilde\mu}\right)  + {\cal O}\left(\ln \ln
\ln\frac{1}{\lambda}\right)\right] \; .
\end{equation}
The important point is that the dependence on boundary conditions
and the initial distribution is solely \textit{logarithmic}, thus
we may expect out predictions to be very robust with respect to
changes of the initial conditions. In particular, the scale factor
at this non-linear time scale is given by
\begin{equation}\label{Cnl} C(\eta_{nl}) = \frac{T_R}{\sqrt{M_* \; T_c}}
\left( \frac{24}{\lambda+3\;e^2} \right)^{\frac14} \;
\sqrt{\ln\frac{1}{\lambda}}\left[1+{\cal
O}\left(\frac{1}{\ln\frac{1}{\lambda}}\right)\right] \; ,
\end{equation}
\noindent where we have used eqs. (\ref{H-T}) and
(\ref{crittemp}).

The amplitude of the long-wavelength modes at the non-linear time,
roughly speaking at the end of the phase transition is
approximately
\begin{equation}\label{endPT}
|f_q(\eta _{nl})| = \sqrt{\frac{32 \; \pi^{\frac52}}{\lambda \;
(1+2\,n_0) \; \tilde\mu }} \;\left[ \ln\frac{32 \;
\pi^{\frac52}}{\lambda \; (1+2\,n_0)}\right]^{\frac14-\frac{q^2}{2
\, {\tilde\mu}^2}}  \; .
\end{equation}
As we will discuss in detail below this non-perturbative scale
will ultimately determine the strength of the magnetic fields
generated during the phase transition.

During the intermediate time regime the equal times correlation
function is approximately \begin{equation}\label{correl} \langle
\Phi^\dagger_{q,a}(\eta )\Phi_{q,b}(\eta )\rangle = \delta_{a,b}
\; |a_q|^2 \; e^{\tilde\mu^2\, \eta^2 } \;
e^{-\left(\frac{q^2}{\tilde\mu^2}+1 \right) \ln[{\tilde\mu}
\eta]}\;.
\end{equation} and its Fourier transform for long wavelenghts is
of the form
\begin{equation}\label{correlation.lenght}
\langle \Phi_a(\vec x,\eta)\Phi_b(\vec 0,\eta)\rangle =
\delta_{a,b} \; |a_0|^2 \; \frac{e^{\tilde\mu^2\,\eta^2
}}{\tilde\mu \,\eta} \;{\tilde\mu}^3 \left[ \frac{\pi}{\ln
{\tilde\mu} \eta}\right]^{\frac32}\; e^{-\frac{\vec
x^2}{\xi^2(\eta )}}
\end{equation}
\noindent which determines the time dependent correlation length
of the scalar field,
\begin{equation}\label{corrlength}
\xi(\eta ) = \frac{2}{\tilde\mu} \;  \sqrt{\ln {\tilde\mu}\;
 \eta } = 2 \, \sqrt{\frac{\ln\left( \sqrt{\mu \; H_R} \, \eta\right)
 }{\mu \;  H_R}} \;.
\end{equation}
This expression is valid in the intermediate time regime
$\eta_c<\eta <\eta_{nl}$ during which the non-equilibrium
dynamics is dominated by the spinodal instabilities. The detailed
analysis of the dynamics in refs.~\cite{nuesfrw,nuestros,destri}
and the discussion of the main features presented above can be
summarized as follows:

\begin{itemize}
\item{At intermediate times $\tilde\mu^{-1}\ll \eta  \leq \eta
_{nl}\sim\tilde\mu^{-1}\sqrt{\ln1/\lambda}$ the mode functions
grow exponentially for modes in the spinodally unstable band
$q<M(\eta)$. The phase of these mode functions \emph{freezes},
namely, becomes independent of time and slowly varying with
momentum.}

\item{ At a time scale determined by the spinodal time the
back-reaction shuts off the instabilities and the phase transition
is almost complete. This can be understood from the following:
the backreaction becomes comparable with the
tree-level term (for $\eta> \eta_R$) when
$\frac{\lambda}{2N}\langle \Phi^\dagger \Phi \rangle \approx
\tilde\mu^4\eta^2$. This relation determines that the mean square
root fluctuation of the scalar field probes the minima of the tree
level potential. }

\item{During the spinodal stage the correlation length of the
scalar field grows in time and is given by eq.(\ref{corrlength}).
This is interpreted as the formation of correlated domains that
grow in time, and is the hallmark of the process of phase
separation and ordering. This correlation length will be important
in the analysis of the correlation of magnetic fields later.  }

\item{The large fluctuations associated with the growth of
spinodally unstable modes of the charged fields will lead to
\emph{current} fluctuations which in turn will lead to the
generation magnetic fields. Thus the most important aspect of the
non-equilibrium dynamics of the charged fields during the phase
transition is that large fluctuations of the charged fields
associated with the spinodal instabilities will lead to the
generation of magnetic fields. Since the modes with longer
wavelength are the most unstable the magnetic field generated
through the process of phase separation will be of long
wavelength. Furthermore we expect that the magnetic field
generated by these non-equilibrium processes will be correlated on
length scales of the same order as that of the charged field
above.  }

\end{itemize}

\subsubsection{Scaling stage}\label{sec:scaling}

A remarkable result of the evolution in the asymptotic regime
(when the effective temperature has vanished) found in
ref.~\cite{scaling} is that there is a very precise cancellation
between the tree level term $-\mu^2 C^2(\eta)$ and the back
reaction $\lambda \Sigma(\eta)$ in the equations of motion
(\ref{modeeqnew}). The self-consistency condition requires that
for a radiation dominated cosmology\cite{scaling}
\begin{equation}\label{conscond}
\lambda\Sigma(\eta)-\mu^2
C^2(\eta)\stackrel{\eta\rightarrow\infty}{=}-\frac{15}{4\eta^2}
\end{equation}
In this asymptotic regime the solutions of the equations of motion
\begin{equation}\label{asyeqn}
\left[\frac{d^2}{d\eta^2}+k^2-\frac{15}{4\eta^2}\right]f_k(\eta)=0
\end{equation}
\noindent are given by
\begin{equation}
f_k(\eta) =  \sqrt{\eta} \;\left[ \; A_k \;\frac{J_2(k\eta)}{k^2}+B_k \;
k^2 \;  N_2(k\eta) \;\right]\; .
\end{equation}
This solution can be written in terms of the scaling variable
\begin{equation}\label{scalevar}
x=k\; \eta
\end{equation}
in a more illuminating form
\begin{equation}
f_k(\eta)=A_k\; \eta^{5/2}\; \frac{J_2(x)}{x^2}+B_k\; \frac{x^2
N_2(x)}{\eta^{3/2}}  \; ,
\end{equation}
As discussed in detail in ref.\cite{scaling}, the relevant
integrals are dominated by $x \sim 1$, namely by modes with
wavelength of the order of the Hubble radius, thus  the second
contribution proportional to $N_2(x)$ can be safely neglected at
long times.

For $x \lesssim 1$ in the long time regime we can further
approximate $A_k \sim A_0$ and the asymptotic solution during this
stage is of the \emph{scaling form}
\begin{equation}\label{scaleform}
f_k(\eta)=A_0\; \eta^{5/2}\; \frac{J_2(x)}{x^2}.
\end{equation}
Since for $x \lesssim 3 $ and large time the modes with small
wavevector have the largest amplitudes, these dominate the
backreaction. The very precise cancellation (\ref{conscond}) leads
to the following sum rule~\cite{scaling}
\begin{equation}\label{sumrule}
\frac{\lambda}{8\pi^2}\int k^2 \; dk \left|f_k(\eta)\right|^2 \;
(1+2 \; n_0) \stackrel{\eta\rightarrow \infty}{=} \tilde{\mu}^4
\; \eta^2
\end{equation}
Since for large $\eta$ the integral is dominated by soft modes
$k\sim\frac{1}{\eta} \rightarrow 0$ the distribution function can
be approximated by $n_0$ and the amplitude by $|A_0|^2$. The sum
rule eq.(\ref{sumrule}) then  leads to the identity
 \begin{equation}\label{A0^2}
|A_0|^2 \; (1+2n_0) =\frac{30\pi^3}{\lambda} \; \mu^2\; H^2_R \; ,
\end{equation}
where we used the integral\cite{gr}
$$
\int_0^\infty \frac{dx}{x^2} \; J_2^2(x)=\frac4{15\pi}\;.
$$
\noindent which is dominated by $x \lesssim 3$.   This is a
remarkable result: the product $|A_0|^2 \; [1+2 \; n_0]$ in the
scaling regime \emph{does not depend on the initial conditions on
the evolution}, namely it is universal in the sense that it is
independent of the previous history through the phase transition.
This is an important result which will play an important role
in the power spectrum of the magnetic fields.

An important consequence of this scaling solution is that the
equal time  two-point correlation function of the scalar field is
given by
\begin{equation}\label{correlation}
\langle \Phi_a(\vec x,\eta)\Phi_b(\vec 0,\eta)\rangle =
\delta_{a,b}~ D(z)~~;~~ z=\frac{|\vec x|}{2\eta}
\end{equation}
\noindent which reveals that the correlation length is given by
the size of the causal horizon\cite{scaling}. The dynamical
evolution during the scaling stage is precisely determined by the
growth of horizon-sized domains\cite{scaling}.

We summarize below the important features of the solutions in the
scaling regime that will be used in the computation of the power
spectrum of the magnetic fields.

\begin{itemize}
\item{For $\eta >> \eta_{nl}$ a scaling regime emerges in which
the mode functions are given by eq. (\ref{scaleform}) with $x=k \;
\eta$. This scaling solution describes  the relaxation of
long-wavelength fluctuations of the charged fields. Again the
phase of these modes \emph{freezes} namely is independent of time.
This is important because this fact will entail that the retarded
self-energy of the transverse photon polarization tensor will give
a subleading contribution to the generation of magnetic fields. }

\item{The sum rule eq.(\ref{sumrule}) constrains the product of
the amplitude times the occupation of the long wavelength scaling
modes to be given by eq.(\ref{A0^2}). }

\item{The scaling solution described above is akin to that found
in \emph{classical} models of formation of topological
defects\cite{turok,durrer}. The scaling regime describes the
evolution of long-wavelength fluctuations and the adjustment of
the spatial correlation length of the scalar field to the Hubble
radius\cite{scaling}. }
\end{itemize}

\subsection{Gauge field dynamics}

In a high temperature plasma a very important aspect that must be
taken into account in the  dynamics of gauge fields is the
electric conductivity, which leads to dissipative processes. As
discussed in \cite{magfiI}, the electric conductivity severely
hinders magnetogenesis, and also introduces the diffusion length
scale which could limit the correlation of the magnetic fields
that are generated.

In Minkowski space in equilibrium the conductivity is obtained
from the imaginary part of the photon polarization and it is
dominated by charged particles of momenta $p\sim T$ in the loop
with exchange of photons of momenta $eT < k \ll
T$~\cite{baym,yaffe}. A careful analysis including Debye
(electric) and dynamical (magnetic) screening via Landau damping
leads to the conclusion that the conductivity is given
by~\cite{baym,yaffe}
\begin{equation}\label{sigmacond}
\sigma = \frac{\mathcal{C} \; N \;  T}{\alpha\ln\frac{1}{\alpha \;
N}}
\end{equation}
\noindent with $N$ the number of charged fields and
$\mathcal{C}\sim \mathcal{O}(1)$.

In an expanding cosmology an in particular during phase
transitions, a more precise assessment of the contributions and
meaning of the conductivity must be provided. As it was discussed
in ref.\cite{magfiI}, the fluctuations of the charged fields
during a phase transition will have very different behavior if the
typical wavevector of these modes is of the order of or smaller
than the symmetry breaking scale or much larger than this scale.

Short wavelength modes, those with typical wave vectors much
\emph{larger than} the symmetry breaking scale are insensitive to
the phase transition and are always in local thermodynamic
equilibrium (LTE). For short wavelength modes deep inside the
horizon, the mode functions are of the free field type $f_q(\eta)
\sim e^{iq\eta}/\sqrt{q}$.

Long wavelength modes, those with wavectors of the order of or
smaller than the symmetry breaking scale undergo critical slowing
down and fall out of equilibrium during the phase transition.
These modes become spinodally unstable during the early stages of
the transition as summarized above and analyzed in detail in refs.
\cite{nuesfrw,nuestros,destri,scaling}.

Thus   the contributions from the charged particle fluctuations to
the photon polarization  must be separated into  two very
different regimes: a) the hard momenta $p \gg |M(\eta)| $
correspond to charge fluctuations that are always in local
thermodynamic equilibrium, b) the soft momenta
$p \ll |M(\eta)|$ fall
out of equilibrium and undergo long-wavelength spinodal
instabilities and enter the scaling regime.

The contribution from hard momenta will lead to a large
\emph{equilibrium} conductivity in the medium, while the
contribution to the polarization from soft momenta will contain
all the non-equilibrium dynamics that lead to the generation of
electromagnetic field fluctuations.

As the instabilities during the phase transition develop, the
fluctuations of the charged fields will generate non-equilibrium
fluctuations in the long-wavelength components of the electric and
magnetic fields with  the ensuing generation of long-wavelength
magnetic fields. However the large conductivity of the medium will
hinder the generation of electromagnetic fluctuations, hence the
conductivity must be fully taken into account to assess the
spectrum of the magnetic and electric fields generated during the
non-equilibrium stage\cite{magfiI}.

We are interested in the generation of long wavelength magnetic
fields, namely $k<<T$ but also $k<< \alpha^2  \; T$, since within
the astrophysical application, the wavelength of interest for
magnetic fields are of galactic scale, while $T$ corresponds to a
wavelength at the peak of the CMB $\sim$ millimeters. Thus the
physical situation corresponds to studying the photon polarization
tensor for long-wavelength.

The polarization tensor has local (tadpole) and non-local
contributions. The equilibrium contribution to the tadpole $\sim
\langle \Phi^\dagger \Phi \rangle$ is dominated by momenta $\sim
T_R$ leading to the hard-thermal loop result $\langle \Phi^\dagger
\Phi \rangle \sim T_R^2$. This contribution is actually cancelled
by the zero frequency-momentum limit of the non-local
polarization (bubble diagram). This cancellation in equilibrium is a
consequence of 
the Ward identities and implies the  vanishing of the magnetic
mass in absence of symmetry breaking\cite{lebellac,scalarQED}. In
equilibrium the inverse propagator vanishes in the limit of zero
frequency and momentum\cite{scalarQED,lebellac} in absence of
symmetry breaking at any temperature. 

Out of equilibrium, the tadpole and the non-local
polarization (bubble diagram) exactly cancel each other in the
longwavelength, longtime limit as shown in ref.\cite{Boyanovsky:1999jh}
[see eq.(6.12) in the reference] indicating the vanishing of the
photon mass.

In equilibrium the long-wavelength and low frequency limit
($k,\omega \rightarrow 0$) of the spatial and temporal Fourier
transform of the transverse polarization is given by
\begin{equation}\label{equi}
\Pi_T(k,\omega)= i \; \omega  \; \sigma
\end{equation}
Thus we write for the full transverse polarization for
long-wavelength electromagnetic fields
\begin{equation}\label{pola}
\Pi_T(\eta,\eta',k)= \sigma \frac{d}{d\eta'} \delta(\eta-\eta')
+\Pi_{noneq}(\eta,\eta',k)
\end{equation}
The first term above includes the contribution from the hard
momentum modes $p \sim T_R$ in the transverse polarization, while
$\Pi_{noneq}(\eta,\eta',k)$ is the contribution from the long
wavelength modes which are unstable in the spinodal
stage and take the scaling form in the scaling regime. Thus in a
very well defined sense, the polarization (\ref{pola}) describes
the effective low energy theory for the transverse photon field.

Our strategy is to obtain the non-equilibrium contribution to the
spectrum of electromagnetic fields to lowest order in $\alpha$ but
treating the conductivity \emph{exactly}.

In a cosmological space-time, the temperature scales with the
inverse of the conformal factor
\begin{equation}\label{T.eta}
T(\eta)=\frac{T_R}{C(\eta)}
\end{equation}
and therefore the conductivity $\sigma=\sigma(\eta)$ becomes
time-dependent. If we are interested in time scales where the
number of ultrarelativistic charge carriers does not change
significantly, which is the case that we will consider in what
follows, then the time evolution of the conductivity is purely
kinematic:
\begin{equation}\label{conductivity.eta}
\sigma(\eta)=\frac{\sigma_R}{C(\eta)}\;.
\end{equation}
An important effect of the conductivity, as discussed in
\cite{magfiI}, is the introduction of a diffusion scale in the
transverse photon propagator. The long-time behavior of the zeroth
order propagators for the transverse gauge fields: retarded (R),
advanced (A), symmetric (H)
$$
\mathcal{D}^{ij}_{R,A,H}(\eta,\eta',k)={\cal P}^{ij}({\hat{\bf
p}}) \; \mathcal{D}_{R,A,H}(\eta,\eta',k)
$$
obey\cite{turnerwidrow} (see\cite{magfiI} for details)
\begin{eqnarray}
&&\left[\frac{d^2}{d\eta^2}+k^2+\sigma(\eta) \; C(\eta) \;
\frac{d}{d\eta}
\right]\mathcal{D}_R(\eta,\eta',k)=\delta(\eta-\eta')~~;
~~\mathcal{D}_R(\eta,\eta')=0~~
\mathrm{for}~\eta<\eta'\cr \cr
&&\left[\frac{d^2}{d\eta^2}+k^2+\sigma (\eta) \; C(\eta) \;
\frac{d}{d\eta}
\right]\mathcal{D}_A(\eta,\eta',k)=\delta(\eta-\eta')~~;
~~\mathcal{D}_A(\eta,\eta')=0~~
\mathrm{for}~\eta>\eta'\cr \cr
&&\left[\frac{d^2}{d\eta^2}+k^2+\sigma(\eta) \; C(\eta) \;
\frac{d}{d\eta} \right]\mathcal{D}_{H}(\eta,\eta',k)=0\; ,
\label{con}
\end{eqnarray}
\noindent with the transverse projector
\begin{equation}
{\cal P}_{ij}({\hat{\bf p}})=\delta^{ij}-{\hat{\bf p}}^i{\hat{\bf
p}}^j  \; .
\end{equation}
Due to eq. (\ref{conductivity.eta}) the comoving conductivity
$\sigma_R=\sigma(\eta) \; C(\eta)$ is an \emph{invariant} quantity
in the regime in which the number of ultrarelativistic charge
carriers is constant. The estimate given by eq.(\ref{sigrat})
clearly indicates that during the radiation dominated era between
reheating and the QCD phase transition, $\sigma_R \; \eta\gg1$.

Then for $k\ll \sigma_R$ (which is certainly fulfilled since the
relevant wavevectors are $k\ll T \ll T/\alpha \sim \sigma_R$) and
$\eta\gg 1/\sigma_R$  we can safely neglect the second order time
derivatives in eqs.(\ref{con}), leading to the
following equations,
\begin{eqnarray}
&&\mathcal{D}_R(\eta,\eta',k)=
\mathcal{D}_C(\eta,\eta';k)\;\theta(\eta-\eta') \quad , \quad
\mathcal{D}_A(\eta,\eta',k)=\mathcal{D}_C(\eta,\eta';k)\;\theta(\eta'-\eta)
\label{advasig} \\
&&\mathcal{D}_H(\eta,\eta',k)=
i\;\frac{e^{-\frac{k^2}{\sigma_R}(\eta+\eta')}}{\sigma_R}\
\label{homosig}\quad , \quad \mathcal{D}_C(\eta,\eta';k)=
\frac{e^{-\frac{k^2}{\sigma_R}(\eta-\eta')}}{\sigma_R}\label{DC}
\; .
\end{eqnarray}

\section{Magnetic field spectrum }
As discussed in detail in reference\cite{magfiI}, the quantity of
astrophysical relevance is the correlation function
\begin{equation}\label{corrfuncB} <\hat B^i(\eta ,\vec x) \hat B^i(\eta ,\vec
0)>_{\rho} \; ,
\end{equation}
where the sum on repeated indices is understood. $B(\eta ,\vec x)$
above is a \emph{Heisenberg operator} and the expectation value is
in the initial density matrix. From this quantity, the spectrum of
the magnetic field is obtained in the coincidence limit
\begin{equation}\label{def.spectrum.B} S_B(\eta ,k)=\frac12\lim_{\eta'\to \eta }\int
d^3x <\{\hat B^i(\eta ,\vec x), \hat B^i(\eta ',\vec 0)\}>_\rho
e^{i\vec k\cdot \vec x}\;, \end{equation} where $\{\;,\;\}$
denotes the anti commutator. And from $S_B(\eta ,k)$ we can
extract the \emph{physical} magnetic energy density  stored on
\emph{comoving} length scales larger than a given $L$
\begin{equation}\label{magn.energy}
\Delta\rho_B(L,\eta )= \frac{1}{2\pi^2 }\int_0^{\frac{2\pi}{L}}
k^2 \; S_B(\eta ,k) \; dk   \; .
\end{equation}
where we have restored the powers of the scale factor arising from
the transformation to conformal time. Denoting by $\Delta
\rho_B(L,\eta)$ the contribution from the non-equilibrium
generation (subtracting the local thermodynamic equilibrium
contribution), a quantity of cosmological relevance to assess the
relative strength of the generated magnetic field is given by the
ratio of the power on scales larger than $L$ to the energy density
in the radiation background
\begin{equation}\label{ratio}
r(L,\eta)= \frac{\Delta\rho_B(L,\eta)}{\rho_{\gamma}(\eta)} \; ,
 \end{equation}
\noindent where \begin{equation}\label{CRB} \rho_\gamma
=\frac{\pi^2 T^4_R}{15 }
\end{equation}
is the \emph{comoving} energy density in the thermal equilibrium
background of photons.

The \emph{physical}  energy densities $ \Delta
\rho_{B,phys}(L,\eta) \; , \rho_{\gamma,phys} $ are obtained from the
comoving expressions above by rescaling $\rho \rightarrow
\rho/C^4(\eta)$ as can be seen from the conformal rescaling
(\ref{physicalfields}). Thus the ratio $r(L,\eta)$ would be a
constant in the absence of non-equilibrium generation or
dissipative processes. Hence  the time dependence of the ratio
(\ref{ratio}) only is solely a consequence of  the non-equilibrium
generation mechanisms or dissipative processes (such as magnetic
diffusion in a conducting plasma) but not through the cosmological
expansion.

Using the results obtained in reference\cite{magfiI} that lead to
a first principle derivation of  the spectrum, we just quote its
expression to leading order in $\alpha$, (here and in what follows
$S_B$ refers solely to the non-equilibrium contribution to the
spectrum)
\begin{equation}\label{S.fund}
S_B(\eta,k) = S^I_B(\eta,k)+ S^H_B(\eta,k)
\end{equation}
\noindent with
\begin{eqnarray}\label{S.B.longexpr}
&&S^I_B(\eta,k)=e^2 N\int \frac{d^3q}{(2\pi)^3}
q^2(1-\cos^2\theta)  \; 
\left[(1+n_q)(1+n_{|\vec q+\vec k|})\left|\int_{\eta_R}^{\eta}
d\eta_1\;k\;\mathcal{D}_C(\eta,\eta_1,k ) f_q(\eta_1)f_{|\vec
q+\vec k|}(\eta_1)\right|^2+\;\right.\nonumber\\
&&\left.+(1+n_q)n_{|\vec q+\vec k|}\left|\int_{\eta_R}^{\eta}
d\eta_1\;k\;\mathcal{D}_C(\eta,\eta_1,k ) f_q(\eta_1)f^*_{|\vec
q+\vec k|}(\eta_1)\right|^2+\;
n_q(1+n_{|\vec q+\vec k|})\left|\int_{\eta_R}^{\eta}
d\eta_1\;k\;\mathcal{D}_C(\eta,\eta_1,k ) f^*_q(\eta_1)f_{|\vec
q+\vec k|}(\eta_1)\right|^2+\;\right.\nonumber\\
&&\left. +n_qn_{|\vec q+\vec k|}\left|\int_{\eta_R}^{\eta}
d\eta_1\;k\;\mathcal{D}_C(\eta,\eta_1,k ) f^*_q(\eta_1)f^*_{|\vec
q+\vec k|}(\eta_1)\right|^2\;\right] \; .
\end{eqnarray}
\noindent and
\begin{equation}\label{Shomo} S^H_B(\eta,k) =
-ik^2F(\eta,\eta;k) \; ,
\end{equation}
\noindent where $F(\eta,\eta';k)$ satisfies the homogeneous
differential equation
\begin{equation}\label{Feqn}
\left[\frac{d^2}{d\eta^2}+k^2+\sigma(\eta)C(\eta) \frac{d}{d\eta}
\right]F(\eta,\eta',k)+\int
d\eta_1\left[\Pi^l(\eta_1)\delta(\eta-\eta_1)+\Pi_R(\eta,\eta_1)
\right]F(\eta_1,\eta',k)=0 \; ,
\end{equation}
\noindent with $\Pi^l(\eta) , \; \Pi_R(\eta,\eta')$ being the one loop
tadpole (local) and retarded (non-local) contributions transverse
polarization\cite{magfiI}.

We note that the function $F(\eta,\eta';k)$ obeys the same
equation as the transverse gauge mean field\cite{magfiI}, but as
it will be argued in detail below, its contribution to the
spectrum of primordial magnetic fields generated during the phase
transitions is subleading in the scalar coupling constant
$\lambda$. The equation of motion (\ref{Feqn}) can be solved
systematically in an expansion in powers of the non-equilibrium
polarization,
\begin{eqnarray}\label{pertF}
&&F(\eta,\eta;k)  =
F^{(0)}(\eta,\eta;k)+F^{(1)}(\eta,\eta;k)+{\cal
 O}(\alpha^2)  \; , \nonumber \\
 &&F^{(0)}(\eta,\eta',k)=\mathcal{D}_H(\eta,\eta',\vec k) \; ,\nonumber \\
 &&F^{(1)}(\eta,\eta',k)=   \int^\eta_{\eta_R}d\eta_1
\mathcal{D}_C(\eta-\eta_1,\vec k) \int^\eta_{\eta_R} d\eta_2
\left[ {\Pi}^{l}(\eta_2)
~\delta(\eta_1-\eta_2)+\Pi_R(\eta_1,\eta_2;\vec k) \right]
\mathcal{D}_H(\eta_2-\eta',\vec k)+ \nonumber \\& & +\quad
(\eta\leftrightarrow \eta') \; ,
\end{eqnarray}
\noindent where $\Pi^l~ \;~ , \Pi_R$ are the tadpole (local) and the
retarded contribution from the one-loop transverse photon
polarization respectively (for details see\cite{magfiI}).

These are given by ~\cite{magfiI}
\begin{eqnarray}\label{leadtad}
&&\Pi^{l}(\eta) = -ie^2 N \int \frac{d^3q}{(2\pi)^3} \;
{G}_>(\eta,\eta,q) \quad  ,  \quad \Pi_{R}(\eta_1,\eta_2,\vec
k)=\left[\Pi^>(\eta_1,\eta_2,\vec k)-\Pi^<(\eta_1,\eta_2,\vec
k)\right]\Theta(\eta_1-\eta_2)
\end{eqnarray}
\noindent with
\begin{eqnarray}
 && \Pi_>(\eta_1,\eta_2, k) =2ie^2 N\int
\frac{d^3q}{(2\pi)^3} \; q^2 \; (1-\cos^2
\theta)~G_>(\eta_1,\eta_2,q) \; G_>(\eta_1,\eta_2,|\vec q+\vec
k|)\; , \; 
\Pi_<(\eta_1,\eta_2, k)=\Pi_>(\eta_2,\eta_1, k)\; .\nonumber
\end{eqnarray}
The scalar propagator $G_>(\eta,\eta';k)$ is constructed from the
mode functions $f_q(\eta)$ that satisfy the mode equations
(\ref{parab.cylinder}) and is given by
\begin{equation}\label{scalpropa}
G_>(\eta_1,\eta_2;k)=\frac{i}{2}
\left[(1+n_k)f_k(\eta_1)f_k^*(\eta_2)+n_k
f_k^*(\eta_1)f_k(\eta_2)\right]\;.
\end{equation}
Therefore
\begin{equation}\label{piretima}
\Pi_{R}(\eta_1,\eta_2,\vec k)= 4e^2 N\int \frac{d^3q}{(2\pi)^3} \;
q^2 \; (1-\cos^2 \theta)~\mbox{Im}\left[G_>(\eta_1,\eta_2,q) \;
G_>(\eta_1,\eta_2,|\vec q+\vec k|)\right]\Theta(\eta_1-\eta_2) \;
.
\end{equation}
This expression for the retarded self-energy must be contrasted
with that of the contribution from $S^I_B(\eta,k)$ which requires
the \emph{real} part $\mbox{Re}\left[G_>(\eta_1,\eta_2,q) \;
G_>(\eta_1,\eta_2,|\vec q+\vec k|)\right]$. This is an important
difference, the long wavelength modes of largest amplitude in
either phase given by (\ref{asimod}) or by (\ref{scaleform}) are
such that their phases are frozen, namely they do not depend on
time, therefore the products $f_q(\eta_1)f_{|{\vec p}+{\vec
q}|}(\eta_2)$ with only the growing mode solutions are
\emph{real} and  such products will contribute only to
$S^I_B(\eta,k)$. This freezing of
phases is a consequence of the classicalization of the scalar
field fluctuations\cite{scaling}.

\bigskip

We now argue that the contribution from $S_B^H$ is subleading.
First of all, the term $F^{(0)}(\eta,\eta;k)$ in eq.(\ref{pertF}) is
the solution of the homogeneous equation in
absence of non-equilibrium fluctuations and leads to the local
thermodynamic equilibrium contribution to the power spectrum,
which is independent of the non-equilibrium generation mechanisms.
This contribution has been analyzed in detail in ref.\cite{magfiI}
and will be subtracted. In what follows we focus solely on the
contribution from the non-equilibrium fluctuations.

For intermediate times after the phase transition during the
\emph{spinodal stage} $\eta_c\leq \eta <\eta_{nl}$,  the
long-wavelength mode functions are approximately given by eq.(\ref{asimod}).

Near the end of the phase transition for $\eta \sim \eta' \sim \eta_{nl}$ the
leading order time dependence of the scalar Green's functions is
approximately given by
\begin{equation}\label{GFts}
f_k(\eta)f_k(\eta') \propto\frac{1}{\lambda} \; .
\end{equation}
where we used eq.(\ref{endPT}).
Thus, the contribution from the tadpole (local term in the
self-energy) is of the order
\begin{equation}\label{tadts}
\Pi^{tad}(\eta)_{\eta\sim \eta_{nl}} \propto \frac{e^2}{\lambda} +
\mathrm{subleading} \; .
\end{equation}
This estimate is consistent with the fact that the tadpole
contribution is $ e^2 <\Phi^{\dagger} \Phi> $ and near the end of the phase
transition the mean square root fluctuations of the scalar field
probe the vacuum state, namely $<|\Phi|^2> \sim \mu^2 /\lambda$.
Since the phases of these modes are frozen, there is no
contribution from the leading order to the retarded polarization,
since it requires the imaginary part of the product of propagators
as displayed in eq. (\ref{piretima}). Because of this cancellation
of the leading term, the contribution from the retarded
polarization bubble is of the same order as that of the
tadpole~\cite{Boyanovsky:1999jh,magfiI}.
\begin{equation}\label{pireta}
\Pi_R \propto \frac{e^2}{\lambda} \; .
\end{equation}
A similar argument based on the sum rule (\ref{sumrule}) leads to
the same conclusion in the scaling regime.

For late times, the $ k \to 0 $ limit of the retarded polarization
exactly cancels the contribution from the tadpole [see eq.(6.12) in
  ref.\cite{Boyanovsky:1999jh}]. 

The contribution from $S^I_B$ is in both cases of ${\cal
O}(1/\lambda^2)$ since  each long-wavelength mode function is of
order $1/\sqrt{\lambda}$ at the end of the spinodal stage or, by
the sum rule (\ref{sumrule}) in the scaling regime. Thus we can
safely neglect the contribution from $S^H_B$ to the magnetic
spectrum.

 \bigskip

Thus the leading contribution to the power spectrum generated by
non-equilibrium fluctuations  is given by
\begin{equation}\label{SButil}
S_B(\eta,k)=(1+2n_0)^2 \;\frac{\alpha N~k^2}{\pi ~\sigma^2_R} \;
e^{-\frac{2k^2}{\sigma_R}\eta} \; \int_0^{\infty} q^4 dq~ d(\cos
\theta) \;
(1-\cos^2\theta)\left|\int_{\eta_R}^{\eta}e^{\frac{k^2}{\sigma_R}\eta_1}\;
f_q(\eta_1)\; f_{|\vec q+\vec k|}(\eta_1)\; d\eta_1\right|^2 \; .
\end{equation}
where $ \theta $ is the angle between the vectors $ \vec q $ and $
\vec k $ and where we have replaced
$$
(1+2n_q)(1+2n_{|\vec q+\vec k|})\simeq (1+2n_0)^2 \; ,
$$
since as highlighted  in section (\ref{sec:spino}) the
 dynamics during both the spinodal stage as well as the scaling stage is
 dominated by the long-wavelength  modes that acquire non-perturbatively large
  amplitudes.

  The final form of the power spectrum generated by the
  non-equilibrium dynamics given by eq.(\ref{SButil}) is the
  basis for the study of primordial magnetogenesis during the
  different stages after the phase transition.

\subsection{Magnetogenesis during the spinodal stage}

The long-wavelength mode functions in the spinodally unstable band
are given by the expression (\ref{asimod}).

The integral over $\eta_1$ for large $\eta$  can be computed
integrating by parts in eq.(\ref{SButil}) as an expansion in
$1/(\tilde{\mu}\eta)^2$. The integral is dominated by the upper
limit, which leads to the cancellation of the exponentials that
contain the conductivity.

The integrals over momenta and angles in eq.(\ref{SButil}) can be
done straightforwardly when the mode functions are given by
eq.(\ref{asimod}). Thus from eq.(\ref{SButil}) whe obtain the
following expression for the spectrum  of magnetic fields
generated by the non-equilibrium fluctuations
\begin{equation}\label{finSBhiT}
S_B(k,\eta\sim \eta_{nl}) = \frac{512 \, \pi^{\frac{9}{2}} \; N
\;\alpha \; k^2 }{\lambda^2\;\sigma^2_R \; {\tilde\mu}^4
\;\xi^5(\eta_{nl})}~ \; e^{-\frac{1}{4}k^2 \xi^2(\eta_{nl})}
\;\left[ 1 + {\cal O}\left(\frac{1}{\ln\frac{1}{\lambda}}\right)
\right]\;.
\end{equation}
where $ \xi(\eta) $ is given by eq.(\ref{corrlength}). In
obtaining this result we used the following
\begin{equation}
\frac{[1+2\,n_0]^2\;|a_0|^4}{\left({\tilde\mu}\eta_{nl}\right)^6}
\; e^{2\left({\tilde\mu}\eta_{nl}\right)^2} = \frac{1024 \;
\pi^5}{\lambda^2 \; {\tilde\mu}^2}
\end{equation}
[see eq.(\ref{spinotime})] and the identities\cite{gr},
\begin{equation}
\int_{-1}^{+1}dx \; (1-x^2)\; e^{-\frac{4 \; q \; k  \;
x}{{\tilde\mu}^2} \; \ln{\tilde\mu}\eta } =
\frac{{\tilde\mu}^6}{16 \; (q \; k \; \ln{\tilde\mu}\eta
)^3}\left\{\frac{4 \; q \; k }{{\tilde\mu}^2} \;
\ln({\tilde\mu}\eta)  \cosh\left[\frac{4 \; q \; k}{{\tilde\mu}^2}
\; \ln{\tilde\mu}\eta  \right] -\sinh \left[\frac{4 \; q \;
k}{{\tilde\mu}^2} \; \ln{\tilde\mu}\eta  \right] \right\}
\end{equation}
and
\begin{equation}
\int_0^{\infty} q \; dq \;  e^{-\xi^2 \; q^2} \left[ \xi^2 \; q \;
k \; \cosh(\xi^2 \; q \; k) - \sinh(\xi^2 \; q \; k) \right] =
\frac{\sqrt\pi}{8}\; k^3 \;\xi \; e^{\frac{1}{4}k^2 \xi^2} \; .
\end{equation}

Notice that the magnetic field  spectrum (\ref{finSBhiT}) is
independent on the amplitude $|a_0|$ and on the initial occupation
$(1+2n_0)^2$. Therefore this result is  quite robust.

This result is  the same as for the Minkowski space-time (see
eq.(7.47) in ref.\cite{magfiI}), except for a multiplicative
factor $ {\tilde\mu}^6 \; \xi^6(\eta_{nl}) \simeq 8 \;
\ln^3\left(\ln\frac{1}{\lambda}\right) $ and  the expression for
the correlation length in the radiation dominated universe
(\ref{corrlength}).

As in Minkowki space-time, the presence of a high conductivity
plasma severely hinders the generation of magnetic fields.
However, a noteworthy aspect is that up to the non-linear time the
magnetic field is still correlated over the size of the scalar
field domains rather than the diffusion length $\xi_{diff} \approx
\sqrt{\eta/\sigma_R}$. The diffusion scale  determines the spatial
size of the region in which magnetic fields are correlated in the
\emph{absence} of non-equilibrium generation. The ratio between
the domain size $\xi(\eta)$ given by eq.(\ref{corrlength}) and the
diffusion length scale $\xi_{diff}(\eta)$ is given by
\begin{equation}\label{scalesratio}
\frac{\xi(\eta_{nl})}{\xi_{diff}(\eta_{nl})} \simeq
\frac{2}{\tilde\mu} \sqrt{\frac{\sigma_R \;
\ln({\tilde\mu}\eta_{nl})}{\eta_{nl}}}\sim
 \left(\frac{M^2_{Pl}\ln\frac{1}{\lambda}}{\mu^2 }
 \right)^{\frac{1}{4}}\gg 1\; . 
\end{equation} 
Where we have used the relations (\ref{H-T}), (\ref{Mstar}),
(\ref{comovingsigma}) and (\ref{spinotime}). Thus an important
conclusion of this study is that the magnetic fields generated via
spinodal decomposition are correlated over regions comparable to
the size of scalar field domains which are \emph{much larger} than
the diffusion scale.

The spectrum for the  electric field  can be obtained from that of
the magnetic field by  simply replacing $k\;\mathcal{D}_C
\rightarrow \dot{\mathcal{D}}_C$. In the soft regime and for time
scales $\frac{1}{\sigma_R}\ll \eta\ll \frac{\sigma_R}{k^2}$ we
have $ \dot{\mathcal{D}}_C \simeq -k^2/\sigma_R^2 $ whereas $k
\,\mathcal{D}_c\simeq k/\sigma_R$. Therefore the electric field
spectrum is suppressed by a factor $k^2/\sigma_R^2$ with respect
to the magnetic field, namely
\begin{equation}\label{hiTelec}
S^{\sigma_R}_E(t,k)=\frac{k^2}{\sigma_R^2} \;
S^{\sigma_R}_B(t,k)\;.
\end{equation}
Thus, in a high temperature plasma with large conductivity the
non-equilibrium processes favor the generation of magnetic photons
instead of electric photons, and again equipartition is not
fulfilled.

The energy density on large scales $\geq L$ again can be computed
in closed form in the limits $L\gg \xi(t_{nl})$ and $L\ll
\xi(t_{nl})$. We find in the first case from
eq.(\ref{magn.energy}),
\begin{equation}\label{specLcond}
\Delta\rho_B(\eta_{nl},L)=\frac{2^{13} \; \pi^{\frac{15}{2}} }{5
\; \lambda^2} \; \frac{N \; \alpha}{[ {\tilde\mu} \;
\xi(\eta_{nl})]^4 \; \sigma_R^2 \; \xi(\eta_{nl}) \;
L}\;\frac{1}{L^4} ~~;~~ L \gg   \xi(t_{nl})\;.
\end{equation}
We find for the opposite case,
\begin{equation}\label{Lchico}
\Delta\rho_B(\eta_{nl},L)=\frac{3 \times 2^{10} \; \pi^3  \; N \;
\alpha}{ \lambda^2 \; {\tilde\mu}^4 \;\sigma_R^2 \;
\xi^{10}(\eta_{nl})} ~~;~~ L \ll   \xi(t_{nl})\;.
\end{equation}

The ratio of the magnetic energy density on scales larger than $L$
at the spinodal time and the total radiation energy given by the
Stefan-Boltzman law $\rho_\gamma=\pi^2 T_{nl}^4/15$ results,
\begin{equation}\label{ratiorhos}
r(\eta_{nl},L)= \frac{\Delta\rho_B(\eta_{nl},L)}{\rho_\gamma} =
\frac{3 \times 2^{13} \; \pi^{\frac{11}{2}} }{\lambda^2} \;
\frac{N \; \alpha}{[ {\tilde\mu} \; \xi(\eta_{nl})]^4 \;
\sigma_R^2 \; \xi(\eta_{nl}) \; L}\;\frac{1}{(L\, T_R)^4} ~~;~~ L
\gg   \xi(t_{nl})\;.
\end{equation}
This result is the same as for the Minkowski space-time
(see eq.(7.53) in ref.\cite{magfiI}), except for a multiplicative
factor $ {\tilde\mu}^6 \; \xi^6(\eta_{nl}) \simeq 8 \;
\ln^3\left(\ln\frac{1}{\lambda}\right) $ and  the expression for
the correlation length in the radiation dominated universe
(\ref{corrlength}).

The factor $(LT_R)^{-4}$ is purely dimensional and is ultimately
the determinining factor for the strength of the generated
magnetic fields on a given scale. These combinations are
\emph{invariant} under cosmological expansion and are determined
by the ratio of the scales of interest today (galactic) to the
thermal wavelength (today) of the cosmic microwave background
radiation at the Wien peak. In particular $LT_R \sim 10^{25}$ for
$L \sim 1~\mbox{Mpc}$(today) [see eq.(\ref{LTR})].

It is clear that the production during this regime is extremely
small, due to the large values of $(LT)^4$ and of the ratio
$\sigma_R^2/\mu^2$. In order to obtain an estimate for the
amplitude of the seed magnetic field, we consider the following
set of parameters: $\lambda=10^{-2},\;\alpha= 10^{-2}, \; \mu=
 10^{14} \; \mbox{GeV}, \; T_R=10^{16} \; \mbox{GeV}$
(corresponding to a critical temperature $T_c= 10^{15}$ GeV). We
then obtain,
\begin{equation}
 r(L=1Mpc)\sim 10^{-157} \;.
\end{equation}
Therefore, the amplitude of the magnetic field generated during
the spinodal stage is completely negligible. This result is
similar to  the result  obtained in Minkowski space-time in
ref.\cite{magfiI} and is  expected on the basis of dimensional
analysis.

\subsection{Magnetogenesis from the scaling regime}

In the scaling regime $\eta >>\eta_{nl}$ the spectrum of the
magnetic field is given by eq.(\ref{SButil}) with the mode functions
in the scaling regime given by eq.(\ref{scaleform}).

The final expression for the leading contribution, given by
eq.(\ref{SButil}) reveals a noteworthy aspect. As we have argued
above, the modes $k$ of astrophysical relevance today, were well
outside the horizon during the radiation dominated era between
reheating and the QCD phase transition. The mode functions
eq.(\ref{SButil}) attain the largest amplitude at long times for
$x=q\eta \leq 2-3$, thus momenta in the polarization loop that are
within the horizon lead to generation of magnetic fields with
long-wavelengths well outside the horizon. This we believe, is an
important mechanism, loop corrections lead to a coupling between
modes inside the horizon with those outside. Thus in this manner,
causal fluctuations can actually lead to the generation of fields
with wavelengths much larger than the horizon.

Since $ k\eta \ll 1$ the power spectrum eq.(\ref{SButil}) takes the
following form using the scaling mode functions eq.(\ref{scaleform}),
\begin{equation}\label{SBesca}
S_B(\eta,k)=(1+2n_0)^2 \;\frac{\alpha N~k^2}{\pi~\sigma^2_R}\;
|A_0|^2 \int_0^{\infty} dq \; \int_{\eta_R}^{\eta} J_2(q\eta_1) \;
\eta_1\; d\eta_1 \int_{\eta_R}^{\eta} J_2(q\eta_2) \;\eta_2 \;
d\eta_2 \; I(q,k,\eta_1,\eta_2) \; ,
\end{equation}
where we set the exponentials equal to unity in eq.(\ref{SButil})
since $  k\eta \ll 1 $ and $ k \ll \sigma_R $ and
$$
I(q,k,\eta_1,\eta_2) \equiv \int_{-1}^{+1}dx \; \frac{1-x^2}{(q^2
+ k^2 - 2kqx)^2} \; J_2(\sqrt{q^2 + k^2 - 2kqx} \; \eta_1) \;
J_2(\sqrt{q^2 + k^2 - 2kqx} \; \eta_2) \; .
$$
Using the summation theorem\cite{gr}
$$
J_2(\sqrt{q^2 + k^2 - 2kqx}\eta) = \frac{4 (q^2 + k^2 - 2kqx)}{q^2
\, k^2 \, \eta^2}\sum_{l=0}^{\infty} (l+2)\;  J_{l+2}(q \, \eta)
\; J_{l+2}(k \, \eta) \; C_l^2(x)
$$
where the $ C_l^2(x) $ are Gegenbauer polynomials. For $  k\eta
\ll 1 $ the $ l = 0 $ terms dominate and we can use the small
argument behaviour of the Bessel functions  $ J_2(k \, \eta) =
\frac18 (k \, \eta)^2 [  1 + {\cal O}(k^2 \, \eta^2)] $. We
finally obtain,
\begin{equation}\label{Ifin}
I(q,k,\eta_1,\eta_2) = \frac{4}{3 \; q^4} \; J_2(q\eta_1) \;
J_2(q\eta_2) \left[ 1 + {\cal O}(k^2 \, \eta^2) \right] \; .
\end{equation}
Inserting eq.(\ref{Ifin}) into eq.(\ref{SBesca}) yields
\begin{equation}\label{SBes2}
S_B(\eta,k)=(1+2n_0)^2 \;\frac{\alpha N~k^2}{3 \;
\pi~\sigma^2_R}\; |A_0|^2 \int_0^{\infty} \frac{dq}{q^4} \left\{
\eta^2 \left[ J_2^2(q\eta) - J_1(q\eta) \; J_3(q\eta)\right] - (
\eta \rightarrow \eta_R ) \right\}^2 \; \left[ 1 + {\cal O}(k^2 \,
\eta^2) \right] \; ,
\end{equation}
where we used  the formula\cite{gr}
\begin{equation}
\int_{0}^{y} z \; J^2_2(\beta z)  \; dz=
\frac{y^2}{2}\left[J^2_2(\beta y)-J_1(\beta y) \; J_3(\beta y)
\right] \; .
\end{equation}
Since $ \eta \gg \eta_R $ we can neglect the terms with $ \eta_R $
 and we find, for $k\ll\eta^{-1} $
\begin{equation}\label{SBesf}
S_B(\eta,k)= {\cal D} \; \frac{\alpha N}{\lambda^2} \;
\frac{k^2}{\sigma^2_R}\; \mu^4 \; H_R^4 \; \eta^7\left[ 1 + {\cal
O}(k^2 \, \eta^2) \right]  \quad; \quad {\cal D}= 48.61\ldots
 \; .
\end{equation}
where we used eq.(\ref{A0^2}) and we computed numerically the
integral
$$
\int_0^\infty \frac{dx}{x^4}\;[J^2_2(x)-J_1(x)J_3(x)]^2 =
0.0005295\ldots
$$
This integral is dominated by the region $x  \geq 1$, namely, by
modes that are inside the horizon. From the estimate
(\ref{produ}), the corrections ${\cal O}\left(k^2
\xi^2_{diff}(\eta)\right)$ are truly negligible between reheating
and the QCD phase transition.

The dependence on the conformal time $\sim \eta^7$ is a direct
 consequence of the scaling form of the solution for the mode
 functions. The strong time dependence is a consequence of the causal
 relaxation of the Goldstone fields, a result of the phase ordering
 kinetics that entails that the size of the domains grow with the horizon.

The spectrum eq.(\ref{SBesf}) exhibits the following important
features:

\begin{itemize}
\item{{\bf i:} The exponential associated with the diffusion
length cancels out, a reflection that the long time behavior of
the integrals above are dominated by the upper limit. Hence the
final result for the spectrum does not feature the exponential
suppression with the diffusion length. }

\item{{\bf ii:} The result for the spectrum only depends on the
initial amplitud $A_0$ and initial occupation number $n_0$ in the
combination $|A_0|^2(1+2n_0)$ which is constrained by the sum rule
eq.(\ref{sumrule}). Hence the final spectrum is \emph{insensitive} to
the initial conditions on the mode functions or occupations, which
in principle carry information of the early history beginning from
the inflationary stage. This is a consequence of the scaling
solution being a fixed point of the dynamics of the scalar
field\cite{scaling,turok,durrer}. }

\item {{\bf iii:} A noteworthy result is that superhorizon
magnetic fields are generated by the non-equilibrium dynamics of
modes inside but near the Hubble radius. This is a consequence of
the polarization loop, wherein the propagators correspond to
momenta $q$ and $|{\vec q}+{\vec k}|$. The momenta $k$
corresponding the wavevector (scale) of the magnetic field is such
that the wavelength is larger than the Hubble radius, but the
momenta $q$ corresponding to the charged scalar field fluctuations
are inside the horizon. The correlation length of the charged
scalar field is of the order of the Hubble radius. Thus acausal,
superhorizon magnetic fields are generated by \emph{loop} effects.
}

\end{itemize}

In order to reveal the enhancement during the scaling regime in a
more transparent manner,  it is convenient to use the relations
(\ref{H-T}),  (\ref{eta.T}), (\ref{crittemp}) and the explicit
expression for the conductivity (\ref{sigmacond}) in the form
\begin{equation}\label{conductivity}
\sigma_R=c[\alpha,N]\; \frac {N~T_R}\alpha \quad , \quad
c[\alpha,N] \equiv \frac{\mathcal{C}}{\ln[\frac{1}{\alpha \, N}]}
\sim {\cal O}(1).
\end{equation}
Then, the  ratio $r(L,\eta)$ for $ L \gg \eta $ is given by
\begin{equation}\label{rreges}
r(\eta,L) = \frac{240\pi~ {\cal D} ~\alpha^3}{N~c^2[\alpha,N] \; [LT_R]^5}
\; \left(\frac{\mu}{\sqrt{\lambda}T(\eta)}\right)^4 \;
\left(\frac{M_*}{T(\eta)}\right)^3 ~.
\end{equation}
where ${\cal D}$ is given in eq. (\ref{SBesf}).  We note that in the
final result (\ref{rreges}) there is no dependence on
the reheating temperature  but only on the scale of symmetry
breaking $\mu$, the temperature at the time $\eta$ and the scalar
and gauge couplings. This is expected since the non-equilibrium
processes begin in earnest after the phase transition,  local
thermal equilibrium prevailed between the time of reheating and
the phase transition.

The dependence on the scalar self coupling $\propto 1/\lambda^2$
is a hallmark of the non-perturbative nature of the growth of
unstable modes and spinodal decomposition, it is ubiquitous in the
non-equilibrium dynamics of phase
transitions\cite{scaling,nuesfrw,nuestros}.

Large scale magnetogenesis is more efficient for large symmetry
breaking scale $\mu$, since the larger the symmetry breaking
scale, the longer  lasts the scaling stage.

Consider for instance the case in which the symmetry breaking
scale $\mu \sim 10^{13}~\mbox{Gev}$ and $\lambda \sim \alpha \sim
10^{-2}$, corresponding to a critical temperature of order of  a
GUT scale $T_c\sim 10^{15} \mbox{GeV}$ and suppose that the scaling
regime lasts until the electroweak phase transition scale, i.e.
$\eta$ is such that $T(\eta)=T_{EW}\sim 10^2~\mbox{GeV}$. Then the
factor
$$
\left(\frac{\mu}{\sqrt{\lambda}T_{EW}}\right)^4
\left(\frac{M_*}{T_{EW}}\right)^3\sim 10^{100}
$$
compensates for the factor $(LT_R)^{-5}$. Taking $N$ and $g_*$ of
the order of $10$ (these values are taken as representative and
they can be changed simply in the final expressions) we can write
the expression for the ratio as
\begin{equation}\label{r-EW}
r(T(\eta),L) \simeq 10^{-34} \left(
\frac{L}{1~\mbox{Mpc}}\right)^{-5} \left(\frac{T_{EW}}{T(\eta)}
\right)^7 \; .
\end{equation}
Therefore \begin{equation}\label{finrat}r(T(\eta),L) \sim \left\{
\begin{array}{l}
  10^{-34}~\mbox{at\,the\,EW\,transition} \\
  10^{-14}~\mbox{at\,the\,QCD\,transition}\; .
\end{array}\right.
\end{equation}
Thus if the scaling regime lasts until a time between the EW and
the QCD phase transitions the amplitude of the large scale
magnetic fields is within the range necessary to be amplified by
some dynamo models. The amplitude of the seed magnetic field
 is strongly dependent on the duration of the scaling
regime. We have only focused on a scaling regime terminating
either at the EW or QCD phase transition since there will surely
be new phenomena associated with these that must be included in
the dynamics of magnetogenesis.

\subsection{Discussion}

\begin{itemize}

\item{{\bf Validity of the approximations:} There are two main
approximations that were used to obtain the results quoted above,
i) the long-wavelength approximation $k\eta \ll 1$ and ii) the
weak coupling approximation. We now provide an estimate of the
reliability of both these approximations to establish the limit of
validity of our results.

\bigskip

{\bf i): Long-wavelength approximation:} In order to
reach our final result for the rate $r(L,\eta)$  we have explicitly used
a series of
approximations which are valid for long wavelengths but whose
validity must be checked before we reach any conclusion regarding
the spectrum at \emph{small} scales. In particular we must address
the limits of applicability of the result
eq.(\ref{rreges}). This result has been obtained by
integrating the magnetic spectrum on scales $0<k<k_{max}$ with
$k_{max}=2\pi/L_{min}$; the formula for the magnetic spectrum was
valid in the limit
\begin{equation}
k_{max} \; \eta_{max}\ll1\;.
\end{equation}
In order to provide an estimate may take for $\eta_{max}$ to be
the (conformal) time at which the EW phase transition occurs,
namely $\eta_{EW}\sim 1  \; \mbox{GeV}^{-1}$. As discussed in the
introduction, we are considering  a situation in which the
magnetic field is considered as a perturbation of a pre-existing
thermal blackbody background. For consistency this requires that,
\begin{equation}
r(L_{min},\eta_{WE})\ll1\;.
\end{equation} This
relation translates in a condition
\begin{equation}
k_{max} \; \eta_{max}\ll\left[\frac{C
N\alpha^3}{24^2c^2}\left(\frac{T_c}{T_{EW}}
\right)^4\left(\frac{T_{EW}}{M_*}\right)^2\right]^{-1/5}\;.
\end{equation}
For $T_c\sim 10^{15} \mbox{GeV}$ this gives $k_{max} \; \eta_{max}\ll
0.0176$ which in turns translate into \begin{equation} L\gg
L_{min}\sim70 \mbox{ fm}\;.
\end{equation} However, this is the comoving length normalized at
the reheating time. In order to convert to the present time, we
have to take in account the redshift
\begin{equation}
z_R=\frac{T_R}{T_0}\simeq 4\times 10^{28}~~\mbox{for}~T_R\sim
10^{15}\mbox{Gev} \;;
\end{equation}
this gives
\begin{equation}
\left. L_{min}\right|_{today}\sim 0.1 \mbox{ pc}\;.
\end{equation}
 Thus, the approximations
invoked are reliable to estimate the amplitude of primordial seeds
on galactic scales or larger, today.

\bigskip

{\bf ii): Weak electromagnetic coupling:} In order to study the
amplitude for much smaller scales the calculations must be done
without the long wavelength approximations  invoked above. In
this case we must expect a breakdown of perturbation theory and
we cannot give a reliable estimate in the present framework.
Furthermore, for scales well inside the Hubble radius,
microphysical processes \emph{not included} in our approximations,
such as scattering between charged fields and between charged  and
gauge fields must be included. These processes will restore
equilibrium between the different fields, if there is a
substantial transfer of power from the charged fluctuations to the
radiation field, this may lead to a change in the equation of
state and the full backreaction on the metric must be included.
At longer time scales the effects of the backreaction of the gauge
fields on the dynamics of the scalar field, as well as the
non-equilibrium contributions to equation of state and the
Friedmann equations must be included self-consistently.
}

\item{{\bf Generation on short distance scales:  } For scales well
inside the horizon during the scaling regime, namely $q\eta \gg
1$, we must  account for causal microscopic processes that tend to
equilibrate the electromagnetic fields generated by the
non-equilibrium processes. In order to understand these processes
we must look at the kinetics of equilibration. The mode functions
for wavevectors well inside the horizon are Minkowski-like, of the
form
$$
f_q(\eta)=
\frac{\alpha_q}{\sqrt{q}} \; e^{-iq\eta}+\frac{\beta_q}{\sqrt{q}} \;
e^{iq\eta} \;  ,
$$
where the coefficients $\alpha_q,\beta_q$ must be determined from
a full numerical evolution. However the constancy of the Wronskian
entails that
$$
|\alpha_q|^2-|\beta_q|^2 =1
$$
which suggests the identification $|\alpha_q| \equiv
1+\mathcal{N}_q\,;\,|\beta_q|\equiv \mathcal{N}_q$,
$\mathcal{N}_q$ is the number of (asymptotic) quanta created
during the time evolution. This form of the asymptotic mode
functions leads to the equipartition between the electric and
magnetic field generation, since spatial and time derivatives are
the same. In turn this entails that we can understand the
generation of electric and magnetic fields by obtaining a
\emph{kinetic} equation for the number  of photons. Such kinetic
equation must necessarily be of the form
$$
\frac{d N_k(\eta)}{d\eta} =
[1+N_k(\eta)]\Gamma^>_k(\eta)-N_k(\eta)\Gamma^<_k(\eta)
$$
which displays the familiar gain minus loss contributions in terms
of the forward and inverse rates. Eventually a steady state will
be reached which will describe a stationary distribution of
photons. The computation of the forward [$\Gamma^>_k(\eta)$] and
inverse [$\Gamma^<_k(\eta)$] require a \emph{detailed} knowledge
of the distribution $\mathcal{N}_q$\cite{Boyanovsky:1999jh} since
these generalized rates are functionals of these occupation
numbers. Clearly such computation lies beyond the scope of this
article and is a task that we will undertake elsewhere. However,
the kinetic equations above will tend to an equilibrated state of
local thermodynamic equilibrium. }

\item{{\bf Effect on the LSS:} It is important to estimate the
effect of the magnetic field on scales corresponding to those of
the last scattering surface, which today are $L_{LSS}\sim
100~\mbox{Mpc}$. From eq. (\ref{finrat}) we see that at the
electroweak temperature $r(T_{EW},L_{LSS}) \sim 10^{-44}$, taking
the fourth root we can provide an estimate of the temperature
fluctuation induced by the primordial magnetic field $\left.\delta
T/T \right|_{LSS} \sim [r(T_{EW},L_{LSS})]^{\frac{1}{4}} \sim
10^{-11}$ which is negligible compared to the CMB temperature
fluctuation at this scale $\sim 10^{-5}$. On the other hand, a
similar estimate at the time of the QCD phase transition gives
$\left.\delta T/T \right|_{LSS} \sim 10^{-6}$ which is marginally
compatible with the current observations. Thus the reliability of
the approximation of weak gauge coupling combined with the effects
on the temperature anisotropy at the last scattering surface seem
to lead us to conclude that \emph{if} a phase transition during a
radiation dominated era occurs near the GUT scale and results in a
scaling stage, our results for primordial magnetogenesis will be
reliable down to the scale of electroweak symmetry breaking. }

\end{itemize}

\section{Conclusions}

In this article we studied large scale primordial magnetogenesis
during a phase transition in the radiation dominated era after
reheating in a model of $N$-charged scalars coupled to an abelian
gauge field. The spectrum of the magnetic field generated during
the non-equilibrium evolution was computed using the formulation
recently introduced in ref.\cite{magfiI}. The dissipative effects
of the conductivity are included by separating the contribution
from hard modes (with momenta of order $T$) to the polarization
tensor of the gauge fields. These modes  are always in local
thermodynamic equilibrium. The non-perturbative, non-equilibrium
dynamics of the scalar field after the phase transition was
studied in the large $N$ limit. The dynamics after the phase
transition features two distinct stages: an early and intermediate
time, spinodal stage, which is dominated by the growth of
long-wavelength fluctuations, followed by a scaling regime during
which the scalar field becomes correlated over horizon-sized
domains. During both regimes, strong non-equilibrium fluctuations
lead to large current-current correlation functions which entail
the generation of magnetic fields. The scaling regime is the most
effective for primordial magnetogenesis since this stage lasts the
longest. During this stage magnetic fields with superhorizon
wavelengths are generated via \emph{loop} effects, the scalar
field momenta in the polarization loop corresponds to wavelengths
of the order of or shorter than the horizon. Thus causal scalar
field fluctuations lead to the generation of magnetic fields on
superhorizon scales. The generation of magnetic field is hindered
by the large conductivity of the plasma and equipartition between
electric and magnetic fields does not hold. The spectrum of the
primordial magnetic field is insensitive to the magnetic diffusion
length which is sub-horizon during the radiation era.

Our final result for the spectrum generated during the scaling
regime is given by eq.(\ref{SBesf}). The  ratio of the energy
density of the magnetic fields on scales larger than $L$ to the
energy density in the cosmic background radiation
$r(L,\eta)=\rho_B(L,\eta)/\rho_{cmb}(L,\eta)$ is given by equation
(\ref{rreges}). For values of $N$, and the gauge
coupling consistent with particle physics models we find that
\begin{equation}\label{r-EW2}
r(T(\eta),L) \simeq 10^{-34} \left(
\frac{L}{1~\mbox{Mpc}}\right)^{-5} \left(\frac{T_{EW}}{T(\eta)}
\right)^7\; .
\end{equation}
Therefore,
\begin{equation}\label{finrat2}r(T(\eta),L) \sim \left\{
\begin{array}{l}
  10^{-34}~\mbox{at\,the\,EW\,transition} \\
  10^{-14}~\mbox{at\,the\,QCD\,transition} \; .
\end{array}\right.
\end{equation}
Therefore, the large scale primordial magnetic fields generated
during the scaling stage after a phase transition may be a
plausible mechanism to generate primordial magnetic fields which
will  be further amplified by the collapse of protogalaxies and by
astrophysical dynamos.

Probably a phase transition at a temperature much larger than the
electroweak leading to a scaling regime lasting until the QCD
phase transition is ruled out by the temperature inhomogeneities
at the last scattering surface. Furthermore the generation of
electromagnetic fields on sub-horizon scales requires a full
kinetic equation that incorporates the microscopic causal
processes that lead to thermalization, the study of these is
beyond the scope of this article.

{\bf Magnetogenesis \emph{after} the QCD phase transition?} The
model that we studied here is assumed to describe the robust
features from the non-equilibrium dynamics of a charged sector
coupled to a (hyper) charge gauge field. GUT's or SUSY
theories may provide the corresponding framework.

However we now argue that precisely the model studied here can
actually describe the non-equilibrium dynamics \emph{after} the
QCD phase transition(s).  After hadronization and chiral symmetry
breaking most of the hadrons produced will be pions, at least this
is the experimental situation in ultrarelativistic heavy ion
collisions. Neglecting the charge form factor (which is justified
for momenta much smaller than the $\rho$ meson mass $m_\rho \sim
770 ~\mbox{Mev}$) the charged pions couple to the electromagnetic
field with minimal coupling. The chiral transition is conjectured
to be in the same universality class as the $O(4)$ linear sigma
model\cite{wilraja}. Thus the model presented in this article, is
the \emph{low energy effective field theory} for the triplet of
pions, two charged and one neutral. Thus we conjecture that the
study in this article \emph{does} describe the generation of
magnetic fields by long-wavelength pions. Therefore the analysis
of this article can apply to magnetogenesis during the
\emph{chiral} phase transition in QCD. While the charged pions
couple to electromagnetism via the minimal coupling in the
long-wavelength limit, the neutral pion couples to the
electromagnetic field through the chiral anomaly $\pi^0
\rightarrow 2\gamma$ and such process will also produce magnetic
fields. We will study the possibility of large scale primordial
magnetogenesis during the chiral phase transition in QCD in a
forthcoming article.

{\bf Acknowledgements:}  The authors thank M. Giovannini, for
useful discussions. D. B. and M. S. thank N.S.F. for support
through grants PHY-9988720 and NSF-INT-9815064.

\end{document}